\documentclass[a4paper]{article}

\usepackage{arxiv}

\usepackage[utf8]{inputenc}
\usepackage{booktabs}
\usepackage{graphicx}
\usepackage{amsmath}
\usepackage{amsfonts}

\usepackage[sort&compress,numbers]{natbib}

\usepackage{booktabs}       
\usepackage{tabularx}       
\usepackage{multicol}       
\usepackage{multirow}       
\usepackage{siunitx}        
\usepackage{pifont}         
\usepackage{xcolor}
\usepackage[colorlinks=true,citecolor=blue,linkcolor=red]{hyperref}
\usepackage{placeins}
\usepackage{blindtext}
\usepackage{doi}
\usepackage{enumitem}

\usepackage[T1]{fontenc}    
\usepackage{url}            
\usepackage{nicefrac}       
\usepackage{microtype}      
\usepackage{fancyhdr}       
\usepackage{subfiles} 

\graphicspath{{images/}}

\immediate\write18{texcount -inc  
-sum main.tex > /tmp/wordcount.tex}

\immediate\write18{texcount -char -freq
 main.tex > /tmp/charcount.tex}

\usepackage{float}
\usepackage{tcolorbox}

\newfloat{ourbox}{thp}{lop}
\floatname{ourbox}{box}

\newtcolorbox[auto counter,
              number within=chapter,
              number freestyle={\noexpand\arabic{\tcbcounter}}
              ]
              {custombox}[2][]{
                fonttitle=\bf,
                coltitle=black,
                colbacktitle=white,
                colback=white,
                boxrule=0.75pt,
                arc=0mm,
                width=155mm,
                title={\vspace{0.25em}box \thetcbcounter. #2\vspace{0.25em}},
                #1
}
\AtBeginEnvironment{custombox}{\centering\small}

\makeatletter
\newcommand{\customlabel}[2]{%
   \protected@write \@auxout {}{\string \newlabel {#1}{{#2}{\thepage}{#2}{#1}{}} }%
   \hypertarget{#1}{}
}
\makeatother

\definecolor{carnelian}{rgb}{0.7, 0.11, 0.11}
\definecolor{cadmiumgreen}{rgb}{0.0, 0.42, 0.24}

\newcolumntype{P}[1]{>{\centering\arraybackslash}p{#1}}
\newcolumntype{R}{>{\raggedleft\arraybackslash}X}
\newcolumntype{C}{>{\centering\arraybackslash}X}

\title{Inductive biases in deep learning models for weather prediction}
\author{
	Jannik Thuemmel\thanks{Correspondence to: jannik.thuemmel@uni-tuebingen.de} \\
	University of T\"ubingen
	\And
	Matthias Karlbauer \\
	University of T\"ubingen
	\And
	Sebastian Otte \\
	University of L\"ubeck
 	\And
	Christiane Zarfl \\
	University of T\"ubingen
 	\And
	Georg Martius \\
	University of T\"ubingen\\
 	\And
	Nicole Ludwig \\
	University of T\"ubingen
  	\And
	Thomas Scholten \\
	University of T\"ubingen
  	\And
	Ulrich Friedrich \\
	Deutscher Wetterdienst, Offenbach
  	\And
	Volker Wulfmeyer \\
	University of Hohenheim, Stuttgart
    \And
	Bedartha Goswami \\
	University of T\"ubingen
	\And
	Martin V. Butz \\
	University of T\"ubingen
 }
\date{}

\begin{document}
\raggedbottom
\maketitle

\begin{abstract}
Deep learning has gained immense popularity in the Earth sciences as it enables us to formulate purely data-driven models of complex Earth system processes.
Deep learning-based weather prediction (DLWP) models have made significant progress in the last few years, achieving forecast skills comparable to established numerical weather prediction models with comparatively lesser computational costs. 
In order to train accurate, reliable, and tractable DLWP models with several millions of parameters, the model design needs to incorporate suitable inductive biases that encode structural assumptions about the data and the modelled processes. 
When chosen appropriately, these biases enable faster learning and better generalisation to unseen data. 
Although inductive biases play a crucial role in successful DLWP models, they are often not stated explicitly and their contribution to model performance remains unclear.
Here, we review and analyse the inductive biases of state-of-the-art DLWP models with respect to five key design elements: data selection, learning objective, loss function, architecture, and optimisation method. 
We identify the most important inductive biases and highlight potential avenues towards more efficient and probabilistic DLWP models.
\end{abstract}

\section{Introduction}\label{sec:introduction}

Accurate weather forecasts play an important role in different aspects of modern society including agriculture, infrastructure, transportation, aviation, and disaster mitigation. 
Ongoing climate change increases the likelihood of extreme weather patterns on all scales  \citep{stott_how_2016, clarke_extreme_2022}. 
This creates immediate demand for new models that can cope with these more erratic and extreme weather patterns and provide reliable forecasts for the public welfare in the next decades.
Until now, operational weather forecast models, which regularly issue predictions from minutes to days to weeks, rely on numerical weather prediction (NWP, box \ref{box:nwp}).
NWP models are constructed by experts from physics, geoscience, and meteorology as well as from data analysis, statistics, and computer science to perform well.

Over the past three decades, NWP models have made significant improvements in providing highly accurate weather forecasts, particularly up to weekly timescales. For instance, forecast skills, which quantify improvements made over a baseline, for a 5-day forecast increased from slightly over 70\% in the mid-nineties to over 90\% in 2015 (c.f. figure 1 in \cite{bauer_quiet_2015}). 
Termed the `quiet revolution' because of its incremental nature, the progress in NWP was made possible mainly by improvements in modelling subgrid processes, data assimilation and resulting model initialisation, and ensemble prediction (box \ref{box:nwp}). 
The gains in these areas crucially depended on concomitant advancements in computational technologies such as the availability of faster and more efficient computing chips which, however, has shown signs of slowing down in recent years,  sparking new discussions of how NWP models can be adapted to the road ahead~\citep{bauer_digital_2021}.

\begin{ourbox}[t]
\begin{custombox}{Numerical Weather Prediction}
\customlabel{box:nwp}{1}
        Numerical Weather Prediction (NWP) is built upon the Navier-Stokes equations governing fluid dynamics with principles of mass and momentum conservation, the idealised gas law, and the first law of thermodynamics \citep{kalnay_atmospheric_2002}. 
        The result is a system of nonlinear, partial differential equations that model the physical processes unfolding in the atmosphere. 
        To generate weather forecasts, given initial conditions that accurately describe the state of the atmosphere, the set of equations are numerically integrated in time on discretised, three-dimensional grid structures. 
        Operational NWP models can cover a wide range of spatial and temporal forecast scales: from local, sub-hourly kilometre-scales, to mid-term, weekly-scales, to global seasonal scales. 
        Two important factors of variation in performance of different NWP models are choices concerning how subgrid processes are modelled and how observational data is assimilated into an initial condition.
        
        \paragraph*{Subgrid physical processes} Regardless of the NWPs spatial and temporal resolution, physical processes below the model resolution will influence the modelled dynamics. 
        These unresolved processes impact model evolution mainly via its thermodynamics (energy transfer) and dynamics (momentum transfer) and, for example, include radiative-transfer, the formation of clouds, deep convection, and land-atmosphere interactions. 
        Typically, such processes are included in the NWP via `parameterisations', which heuristically quantify their interaction with the properly resolved physical processes. 
        Accurate parameterisations have a strong impact on the predictive power of NWPs as they influence crucial model variables such as precipitation, temperature, and wind \citep{stensrud_parameterization_2007} and better model parameterisations form a central objective of research in Earth System Modelling \citep{schneider_earth_2017}. 
        
        \paragraph*{Ensemble modelling}
        Inherent nonlinearities of atmospheric dynamics impose a hard limit on how far ahead in time we can predict the weather. 
        Small differences in specifying the initial atmospheric state can result in a predicted state that is drastically different from what is observed \citep{slingo_uncertainty_2011}. Probabilistic weather forecasts based on an ensemble of initial conditions, combined with online data assimilation try to tackle the problem by sampling the possible states in the future as best as possible with a given set of initial conditions, and then integrating the model with only those trajectories that are the most probable, given the latest observations \citep{bannister_review_2017}.
        This results in a forecast cycle, which goes from initial conditions to a first ensemble estimate, followed by a re-initialisation based on data assimilation, a second, more constrained ensemble, and so on. 
        Strategies for model initialisation, data assimilation, and ensemble analysis crucially influence the predictive skill of NWP models.
\end{custombox}
\end{ourbox}

Purely data-driven weather forecast models powered by deep learning (DL, box \ref{box:dl}) are on the verge of triggering a major breakthrough in modelling atmospheric dynamics from short-to-medium range weather to climate scale simulations \citep{ben-bouallegue_rise_2023}.
An early study in this direction used convolutional Long Short Term Memory (LSTM) networks for precipitation nowcasting~\citep{shi_convolutional_2015}. 
Initial attempts to model global atmospheric states include a multilayer perceptron~\citep{dueben_challenges_2018} and a convolutional neural network (CNN) architecture (with an LSTM variant) to forecast mid-tropospheric geopotential height~\citep{weyn_can_2019}, as well as an autoencoder setup to forecast the weather within a simple climate model~\citep{scher_toward_2018}. 
Many similar studies that followed over the last four years have demonstrated the potential of using deep learning models for weather prediction (DLWP).

One fundamental reason why it is attractive to use DL for predicting the weather is its apparent simplicity---we do not need to fully understand nor explicitly model the physics of the atmosphere as thoroughly as in NWP.
NWP models themselves are highly complicated objects with many interdependent handcrafted modules and tuning parameters, which make them hard to adapt to new situations.
In contrast, DLWPs may learn the necessary abstract representations of atmospheric dynamics required to predict the weather accurately, and infer the influences of subgrid processes and interactions between them from their training data (box \ref{box:dl}).
Deep learning thus enables faster model adaptation and more flexible inferences of complex correlation patterns as long as sufficient amounts of data are available.
Moreover, once a DLWP model has been trained, the forward inference of a forecast can be conducted at a fraction of the energy and time investment of an NWP forecast \citep{kurth_fourcastnet_2022}.


\begin{ourbox}
\begin{custombox}{Deep Learning}
\customlabel{box:dl}{2}
        Deep learning (DL) refers to artificial neural networks that contain many layers, thus transforming input data via a cascade of abstract, compositional, and hierarchical representations into target predictions \citep{lecun_deep_2015}.
        Each layer in a neural network architecture is parameterised by computational units called neurons or weights.
        These units combine information from previous layers, for example, via a weighted sum, and produce a consequent output with a non-linear but differentiable activation function.
        During training, the weights are tuned via error backpropagation by minimising a task-specific performance criterion.
        The trained model can then be used with its learned set of weights to efficiently generate predictions about new data instances.

        \paragraph*{Latent representations} 
        At their core all DL methods learn projections of data into latent spaces, thereby representing data in ways that are better-suited for the generation of accurate predictions \citep{bengio_representation_2014}.
        This is necessary since observable data spaces are often inadequate to predict future system states efficiently---crucial system dynamics are warped in the data dimensions, take place on a manifold of smaller dimensionality, or are controlled by latent structures and patterns. 
        In a neural network the data manifold gets incrementally transformed from layer to layer, resulting in more abstract representations that are dependent on the training dataset as a whole, the definition of the learning task and the characteristics of individual transformations in the cascade (c.f. box \ref{box:nnp}).

        \paragraph*{Inductive bias} 
        Problem-targeted learning in DL is accomplished by implementing well-motivated structural assumptions about the data and the modelled processes \citep{wolpert_no_1997}, known as inductive biases \citep{battaglia_relational_2018}. 
        Inductive biases can accelerate model convergence and can improve both computational efficiency and generalisation to unseen data. 
        For example, a convolutional DL model induces the bias that pixels close to each other contain correlated information, thus excelling at processing images when compared to a standard artificial network which is unaware about spatial proximity. 
        In principle, any design choice in the learning pipeline creates implicit inductive biases and thus affects model performance.
\end{custombox}
\end{ourbox}

Setting up a DL model, however, requires many design choices, all of which induce implicit and nontrivial  learning and information processing biases. 
Inductive biases (box \ref{box:dl}) essentially implement prior assumptions about the modelled system dynamics, aiming at both keeping the learning problem tractable and fostering generalisation.
In order to develop a DLWP model that is tractable, generalisable to out-of-data scenarios, and explainable, the inductive biases in the DLWP model need to be well-chosen.
We thus aim to conceptualise these design choices and the applied techniques therein to advance our understanding of neural network design in general and to facilitate application and further development of models for DLWP.
For this review we select four high-performing models for short- to mid-range weather forecasting that are representative of state-of-the-art design philosophies: CubedSphereNet, FourCastNet, GraphCast and PanGu-Weather.

\textbf{CubedSphereNet} \citep{weyn_sub-seasonal_2021} and related works \citep{karlbauer2023advancing, weyn_improving_2020, weyn_can_2019} are purely convolution-based neural networks (CNN), utilising geometric preprocessing to better capture global atmospheric dynamics.
The first DLWP model to rival the performance of NWP was \textbf{FourCastNet} \citep{pathak_fourcastnet_2022} which used Fourier Neural Operators to efficiently process high-resolution atmospheric fields---we discuss FourCastNet jointly with its more recent formulation using spherical harmonics \citep{bonev_spherical_2023}.
Around the same time as FourCastNet, Keisler released his DLWP model based on Graph Neural Networks (GNN, \cite{keisler_forecasting_2022}) which heralded the utility of spherical-meshes that would quickly afterwards be used in \textbf{GraphCast} \citep{lam_graphcast_2022} to outperform operational NWP on several metrics.
In concurrent work, \textbf{PanGu-Weather} \citep{bi_pangu-weather_2022} exploited the scaling capabilities of Transformer models to achieve similarly impressive metrics---we further note subsequent work on Transformer models such as FuXi \citep{chen_fuxi_2023}, FengWu \citep{chen_fengwu_2023} and Stormer \citep{nguyen_scaling_2023}. 

While the landscape of DLWP is in a constant flux of rapid development, the way these models encode inductive biases and the structural assumptions they make about DLWP are applicable to older and newer models alike.
In order to identify and characterise distinct inductive biases in these models, it is useful to conceptualise them along the following five design elements: 

\begin{enumerate}
    \item \textbf{Data selection} covers the choice of variables as well as spatial and temporal resolution of the model [section \ref{sec:data_selection}].
    \item \textbf{Learning objectives} address the high level functionality of the model by defining the forecast mode and, potentially, how to handle uncertainty [section \ref{sec:learning_objectives}].
    \item \textbf{Loss functions} are used to quantify how well the model predictions fit the data and enable gradient-based optimisation [section \ref{sec:loss_function}].
    \item \textbf{Neural network architectures} are designed around the computational primitives for data encoding, processing and decoding [section \ref{sec:architectures}].
    \item \textbf{Training} of model parameters defines the learning process by choosing schemes for mini-batching, learning-rate adaptation and curricula of objectives [section \ref{sec:training}].
\end{enumerate}


\begin{table}[htb!]
    \label{tab:data_table}
    \centering
    \scriptsize
    \begin{tabularx}{\textwidth}{lXXrlcc}
        \toprule
        \multirow{3}{*}{\textbf{Model}}     &       \multicolumn{3}{c}{\textbf{Variables}}       &                                \multicolumn{3}{c}{\textbf{Forecast resolution}}                          \\
        \cmidrule(r){2-4}\cmidrule(l){5-7}
        & \multirow{2}{*}{\textbf{Prescribed}} & \multicolumn{2}{c}{\textbf{Prognostic}}         &  \multirow{2}{*}{\textbf{Vertical}} & \multirow{2}{*}{\textbf{Horizontal}} & \multirow{2}{*}{\textbf{Temporal}} \\
         \cmidrule(r){3-4}
         &                                  & \textbf{Surface}          & \textbf{Atmospheric} &                                       &                                                               \\
        \midrule
        \multirow{4}{*}{\textbf{CubedSphereNet}} 
                & Topographic height        & \SI{2}{m} temperature             & Temperature                                           & \SI{850}{hPa}                 & \multirow{4}{*}{$\SI{1.4}{^\circ}$}   & \multirow{4}{*}{6h}   \\
                & Land-sea mask             &                                   & Geopotential                                          & 500, \SI{1000}{hPa}           &                                       &                       \\
                & Incident solar radiation  &                                   & $300 - $\SI{700}{hPa} thickness$^{*}$                 &         --                    &                                       &                       \\
                &                           &                                   & Total column water vapor$^{*}$                        &         --                    &                                       &                       \\
        \midrule
        \multirow{6}{*}{\textbf{FourCastNet}} 
                & \multirow{5}{*}{None}     & \SI{2}{m} temperature     &                       &                                       & \multirow{6}{*}{$\SI{0.25}{^\circ}$}  & \multirow{6}{*}{6h}   \\
                &                           & \SI{10}{m} $U$, $V$ winds & Geopotential          & 50, 500, 850, \SI{1000}{hPa}          &                                       &                       \\
                &                           & Sea level pressure        & $U$, $V$ winds        & 500, 850, \SI{1000}{hPa}              &                                       &                       \\
                &                           & Surface pressure          & Relative humidity     & 500, \SI{850}{hPa}                    &                                       &                       \\
                &                           &                           & Temperature           & 500, \SI{850}{hPa}                    &                                       &                       \\
                &                           &                           & Total column water vapor$^{*}$  &                             &                                       &                       \\
        \midrule
        \multirow{6}{*}{\textbf{GraphCast}} 
                & Latitude and longitude    &                           &                       & \multirow{6}{*}{37 pressure levels}   & \multirow{6}{*}{$\SI{0.25}{^\circ}$}  & \multirow{6}{*}{6h}   \\
                & Land-sea mask             & \SI{2}{m} temperature     & Geopotential          &                                       &                                       &                       \\
                & Geopotential at surface   & \SI{10}{m} $U$, $V$ winds & $U$, $V$, $W$ winds   &                                       &                                       &                       \\
                & Local time of day         & Sea level pressure        & Specific humidity     &                                       &                                       &                       \\
                & Elapsed time of year      & Total precipitation       & Temperature           &                                       &                                       &                       \\
                & Incident solar radiation  &                           &                       &                                       &                                       &                       \\
        \midrule
        \multirow{4}{*}{\textbf{PanGu-Weather}} 
                & \multirow{4}{*}{None}     & \SI{2}{m} temperature     & Geopotential          & \multirow{4}{*}{13 pressure levels}   & \multirow{4}{*}{$\SI{0.25}{^\circ}$}  & \multirow{4}{*}{1h}   \\
                &                           & \SI{10}{m} $U$, $V$ winds & $U$, $V$ winds        &                                       &                                       &                       \\
                &                           & Sea level pressure        & Specific humidity     &                                       &                                       &                       \\
                &                           &                           & Temperature           &                                       &                                       &                       \\
        \bottomrule
    \end{tabularx}
    \begin{flushleft}
        $^{*}$ $300 - $\SI{700}{hPa} thickness and total column water vapor are integrated variables which summarise information about temperature (in case of geopotential thickness) and specific humidity (in case of total column water vapor) across multiple pressure levels, i.e. across different heights in the atmosphere.\\[1.5em]
    \end{flushleft}        
    \normalsize
    \caption{Data selection across considered DLWP models. 
    The choice of \textbf{variables} defines the information the model has access to. 
    Prescribed fields do not depend on model outputs but are known in advance e.g. topographic height or the land-sea mask. 
    Prognostic variables are forecast by the model and evaluated against the loss function.
    The \textbf{forecast resolution} has strong implications for the computational complexity of the learning problem, as well as for the complexity and types of physical processes that can be modelled. 
    We differentiate between vertical resolution, defined on isopotential pressure levels, horizontal resolution defined on a latitude-longitude grid and temporal resolution referring to the model timestep.
    }
\end{table}

\section{Key design elements}
\label{sec:review}

\subsection{Data selection}\label{sec:data_selection}

The major catalyst for the success of deep learning in weather prediction has been the availability of large amounts of high quality data, which enables the training of vastly overparameterised models without risk of overfitting.
The ERA5 dataset, which was used to train all models discussed in this review, offers global atmospheric reanalysis data at an hourly resolution on a $0.25^{\circ}\times0.25^{\circ}$ latitude-longitude grid dating back to 1940 \citep{hersbach_era5_2020}.
Reanalysis provides a compromise between observational data and model simulations by utilising a high quality NWP model to estimate global fields---these fields are tuned to minimise disagreement between available observations and model trajectories.
As a consequence, DLWP benefits immensely from the advancement of NWP that enabled the creation of ERA5 and future generations of reanalysis products.
DLWP model development was further facilitated by the WeatherBench benchmark and its successor WeatherBench2 that established DL-ready preprocessing and evaluation protocols for ERA5 data \citep{rasp_weatherbench_2020, rasp_weatherbench_2023}.

On the global scale atmospheric dynamics evolve as a complex interplay of multiple variables governed by the primitive equations of the atmosphere (c.f. box \ref{box:nwp}).
These equations are defined in terms of temperature, pressure (also called geopotential), humidity, and wind fields in three dimensions (U, V, and W for zonal/eastward, meridional/northward, and vertical components, respectively).
Forecasts are fundamentally about weather impacts on the surface of the earth where they affect human livelihood and infrastructure, as such, surface temperatures, precipitation or near-surface winds are explicitly included in DLWP models.
We emphasise that the interactions with the surface also affects the evolving atmospheric dynamics in non-trivial and important ways and including them likely benefits the accurate prediction of the atmospheric evolution.
Beyond the variables that directly depend on the atmospheric state, it can be beneficial to include prescribed boundary condition information.
For example, geospatial- and temporal coordinates, the land-sea mask or a digital elevation model may enable the model to learn region-specific dynamics and atmosphere-surface interactions, whereas the incident solar radiation can be used to represent the influx of energy from the sun and thus acts as a proxy for the seasonal- and day-and-night cycles.

When we inspect the size of the atmospheric state propagated forward in time by the DLWP models discussed here (table \ref{tab:data_table}) we notice substantial differences.
CubedSphereNet operates on a lower horizontal resolution than the others, heavily reducing the computational requirements, but restricts itself to largely model temperature and geopotential---variables that evolve on larger spatial scales and thus require less resolution.
Information about several pressure levels is provided to the model in the form of vertically integrated variables, further reducing the computational load of the model.
A similar approach is taken in FourCastNet, although later versions use much larger atmospheric states, with the high horizontal resolution being traded for a rather coarse vertical resolution.
GraphCast uses a substantially larger atmospheric state, totalling 227 prognostic variables as well as additional prescribed information. 
This large atmospheric state comes at a substantial cost, but is still much smaller than the states propagated by the reference NWP model.
PanGu-Weather follows the conventions set by WeatherBench to use 13 pressure levels and, in contrast to all other models, a 1-hour timestep.

The resolution at which the continuous atmospheric state should be discretised is a fundamental question in weather and climate modelling and DLWP, too, has to face this question of balancing computational requirements with sufficient fidelity.
As noted in the work of \cite{keisler_forecasting_2022} larger atmospheric states provide a denser, and thus more regularised, target for the model to capture, but as we can see from the CubedSphereNet reduced state sizes are also capable of producing skillful and stable forecasts.
The ideal trade-off likely depends on the impact variable one is most interested and, as always in DL, the available computational resources.

\begin{figure}
    \centering
    \includegraphics[width=\textwidth]{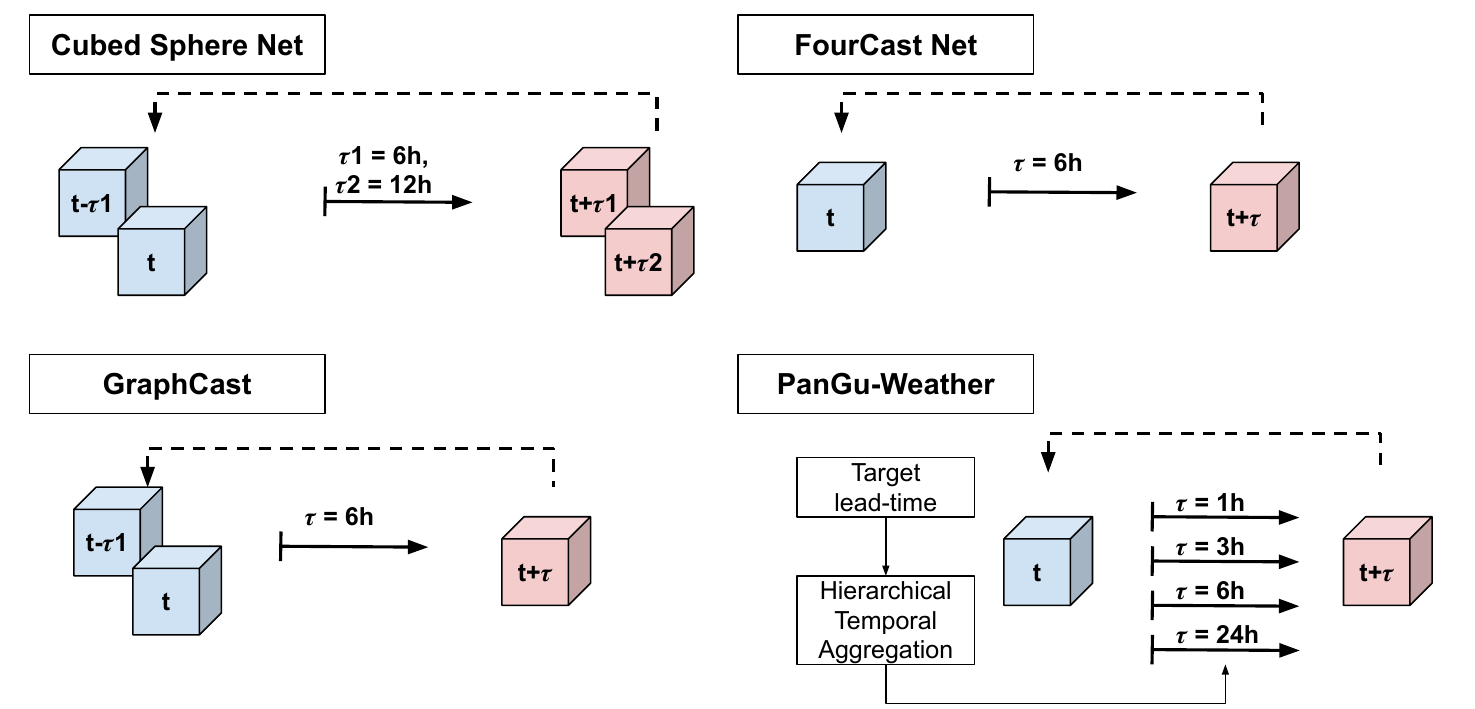}
    \caption{Objective definitions in the four reviewed models. All models shown here support autoregressive roll-outs of, in principle, arbitrary length.
    GraphCast learns a map from two observed states, 6 hours apart, to a subsequent state 6 hours in the future. 
    PanGu-Weather maps a single observed state to a desired lead-time state with intermediate step-sizes being combined from 1, 3, 6 or 24 hours according to a greedy algorithm that minimises the number of autoregressive steps.
    FourCastNet maps a single state 6 hours into the future.
    CubedSphereNet maps two observed states onto the two subsequent states, with each state being 6 hours apart.
    }
    \label{fig:objectives}
\end{figure}

\subsection{Learning objectives}\label{sec:learning_objectives}

Learning objectives define the functional mapping that the model will represent, e.g., a classifier, a predictor, or a generative model.
In the context of DLWP, we discuss objectives which define a (distribution over) future atmospheric state(s) as target and thus model a forecasting task, illustrated in figure \ref{fig:objectives}.
Forecasting objectives depend on the temporal process that is modelled and we follow the distinction between iterative and direct forecasts proposed by \citeauthor{rasp_data-driven_2021}. 
Moreover, we review the challenges involved in generating probabilistic forecasts---differentiating between models that produce uncertainty estimates in deterministic fashion and generative models that can be sampled from to obtain different realisations of the modelled process.

\subsubsection{Forecast models}\label{sec:iterative_forecasts}

\paragraph{Iterative prediction}
Iterative models predict a sequence of future states by feeding their own outputs back as input, forming a closed loop.
We further distinguish two kinds of iterative objectives: auto-regressive and recurrent models. 
Autoregressive models only depend on the last time step to generate a prediction, which turns into the only information available in the next step.
Recurrent models, which mimic hidden Markov models, maintain an internal state representation that allows for memory effects beyond the last frame. 
They are generally able to model more complex temporal interactions at a higher computational cost and more challenging training dynamics \citep{metz_gradients_2022, mikhaeil_difficulty_2022}.

Two issues arise with iterative forecasting methods: (i) a mismatch between ground truth and model generated inputs may lead to quickly divergent predictions, and (ii) temporal patterns that extend beyond the number of steps that are included in the training objective may be fully ignored by the model.
Both issues can in principle be addressed by training the model not only on one-step-ahead predictions, but on longer closed-loop predictions, where the model is forced to deal with its own inputs and subsequent errors \citep{bengio_scheduled_2015}.
However, each additional prediction step linearly increases the computational complexity of the training and inference, necessitating a considerable trade-off.
To achieve ideal trade-offs, curriculum learning strategies (c.f. section \ref{sec:training}) may be applied to dynamically adapt the objective throughout the optimisation procedure.

In DLWP, both kinds of iterative approaches find application, depending on the desired forecast horizon.
Essentially all models that are trained on high horizontal resolutions currently employ an autoregressive setup, due to the otherwise prohibitive computational costs during training. 
Among the models reviewed here, FourCastNet, GraphCast and CubedSphereNet differ in the number of observed and forecast weather states during autoregressive prediction: FourCastNet maps a single state onto the next, GraphCast maps two consecutive states onto the next and CubedSphereNet maps two states onto two subsequent ones.
Recurrent forecasting models have been shown to exhibit preferable skill at longer leadtimes, at least in works that used a reduced horizontal resolution e.g. \cite{hu_swinvrnn_2022, chen_swinrdm_2023} and are especially prevalent in precipitation nowcasting \citep{ravuri_skillful_2021, leinonen_stochastic_2021, adewoyin_tru-net_2021}.

\paragraph{Direct prediction}

Direct forecast models generate a prediction for a particular target lead-time, without any intermediate predictions.
The lack of intermediate predictions eliminates amplifying or vanishing gradient problems, emphasises those temporal features that are most salient for the targeted time scale and is computationally more efficient.
However, generalising beyond the trained time scale can be challenging and if predictions for multiple lead-times are desirable, as is the case in medium-range weather prediction, multiple models have to be trained for each target lead-time.

Alternatively, direct forecasts at different lead-times, termed continuous forecasts, are possible if the target lead-time is included as conditioning information---either in the form of an input field \citep{rasp_data-driven_2021, sonderby_metnet_2020} or as input to a separate conditioning network, as in the work of \cite{espeholt_skillful_2021} and Stormer.
This approach is computationally efficient compared to training multiple direct models for different lead-times, because it allows sharing features that are useful at all forecast time-scales.

PanGu-Weather, can be seen as a combination of iterative and direct forecasting---it utilises an ensemble of one-stepep prediction models, one for 1, 3, 6 and 24 hours respectively, which are greedily combined to generate the shortest autoregressive rollout for a targeted lead-time.
In Stormer the use of learned lead-time conditioning allowed the authors to generate an ensemble of predictions based on different combinations of lead-time prediction chains, rather than the greedy strategy employed by PanGu-Weather.
It is important to note that there are no guarantees that consecutive predictions are temporally consistent, which led the authors of the FuXi models, to take an even more hybridised approach with a cascade of three iterative forecasting models specialised for 1-5, 5-10 and 10-15 day forecasts, respectively.

\subsubsection{Generative models}
Quantifying the uncertainty in a forecast is essential in all weather prediction tasks, especially for long forecast horizons \cite{palmer_decisions_2014}.
While NWPs typically capture model uncertainty indirectly via random perturbations of initial conditions \citep{palmer_ecmwf_2019} and physical paremetrisations, DLWPs can capture model uncertainty in many different ways, incorporating predictive distributions and stochastic sampling procedures in their objectives.
DL models can be trained to directly output a predictive distribution, either as a distribution over quantiles of a variables as in the MetNet models \citep{espeholt_skillful_2021, sonderby_metnet_2020, andrychowicz_deep_2023}, or as parametric (often Gaussian) distribution as in FengWu or an ensemble as in AtmoRep \citep{lessig_atmorep_2023}.

While predictive distributions offer an efficient way to obtain uncertainty estimates, it can often be desirable to be able to sample potential trajectories of weather dynamics from a distribution.
Generative probabilistic models achieve this goal by sampling from a simple distribution (e.g., a standardised Gaussian) and applying a learned transformation on these samples, thereby producing outputs with desired statistics \citep{bond-taylor_deep_2021}.
Different methodologies have emerged on how to represent these transformations with neural networks and train the resulting architectures.

Variational Auto-Encoders (VAE \citep{rezende_stochastic_2014, kingma_auto-encoding_2014}) utilise the so-called reparametrisation trick \citep{kingma_auto-encoding_2014} to learn the parameters of a latent space distribution.
The learning objective of VAE combines reconstruction of the original input with a Kullback-Leibler (KL) divergence that regularises the latent space distribution to contain only as much information as needed.
In \cite{hu_swinvrnn_2022} and \cite{chen_fuxi-s2s_2023} the VAE formulation has been used to learn a data-informed distribution of perturbations to the unfolding dynamics \citep{babaeizadeh_stochastic_2018, denton_stochastic_2018}.
This methodology was found to enable better calibrated ensemble forecasts and longer stable rollouts than what has been observed in non-generative methods.

Generative Adversarial Networks (GAN \citep{goodfellow_generative_2014}) consist of two networks, a Generator that maps from a low-dimensional noise distribution to the distribution of data samples and a Discriminator that tries to classify samples as coming from the original, or the generated data distribution.
The training objective of GANs is defined as an adversarial game between the Generator and the Discriminator, each trying to 'fool' the other as much as possible.
Although GANs have not been successfully used in medium range DLWP, they have been effective in forecasting short-term precipitation \citep{ravuri_skillful_2021} and statistical downscaling of different weather fields \citep{leinonen_stochastic_2021, gong_temperature_2022, hoffmann_atmodist_2022}.

Diffusion models \citep{ho_denoising_2020, sohl-dickstein_deep_2015, song_denoising_2021} are the most prevalent class of generative models in current DL practice.
The objective of diffusion models is to predict noise that has been applied to an input and thereby denoising and reconstructing the unperturbed input.
Once the denoising model has been trained, samples can be generated by directly sampling noise and applying the denoising network iteratively, creating high fidelity samples from the distribution of un-noised inputs.
Works such as GenCast \citep{price_gencast_2023}, DYffusion \citep{cachay_dyffusion_2023} and others \citep{li_seeds_2023, chen_swinrdm_2023, leinonen_latent_2023} have explored the potential of Diffusion models for weather and climate forecasting.
Although their formulations can differ substantially, all of these models essentially treat the evolution of the weather dynamics as the process that induces noise in their observations and subsequently train networks that model the spatiotemporal variations in the data to denoise the observations.

\begin{figure}
    \centering
    \includegraphics[width=\textwidth]{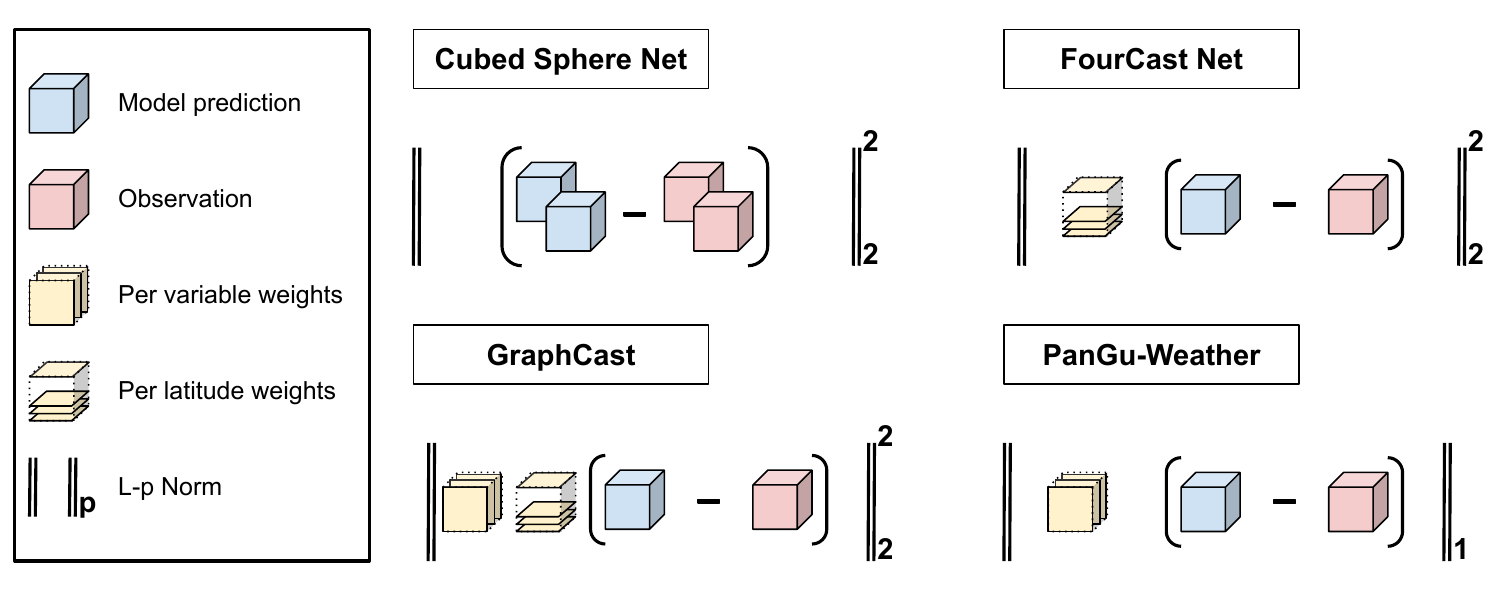}
    \caption{Loss functions used in the four models under consideration. All models utilise an L-p norm between the observed and predicted atmospheric state, evaluated at each spatio-temporal location individually and then averaged across all locations. 
    GraphCast additionally weighs the loss with pre-defined weights per variable and per latitude.
    PanGu-Weather utilises per-variable weights derived from the performance of an earlier training run.
    FourCastNet weighs the loss per latitude and CubedSphereNet does not report any further weighting of the loss.}
    \label{fig:losses}
\end{figure}

\subsection{Loss function}\label{sec:loss_function}

Once the objective is defined, a differentiable loss function measures how well the model predictions match the target and thereby enables subsequent optimisation of the model parameters during training.
figure~\ref{fig:losses} illustrates the losses applied in the reviewed works.
In practice, the quality of the model predictions is often verified with several metrics and additionally with case studies. 
Hence, it may be desirable to train the model with a loss function that reflects the most salient, task-specific, properties for verification.
First, we discuss standard loss functions for regression tasks and, second, highlight the potential of probabilistic loss functions. 

\subsubsection{Deterministic loss functions}
When predicting atmospheric fields, the targets usually are continuous variables distributed over space and time. 
The standard approach to calculate the loss with respect to the objective is to discretise the target values in space and time via a spatiotemporal grid or mesh. 
Losses then typically target minimising a distance between the predicted and true data values, averaged over discretised space and time.
Standard distance measures on real numbers are the L1- and L2-norm, that is, mean-absolute error (MAE) and mean-squared error (MSE), respectively.
However, there does not appear to be a large difference resulting from the choice of the loss, with larger differences being induced by the temporal extent of the loss computation and variable-specific normalisations that account for the differences in variable scales.  

An important insight in the design of DLWP loss functions has been that variable scales depend on geographical location and pressure level and if these differences are not normalised, the loss is dominated by the variables with the largest magnitudes.
\cite{keisler_forecasting_2022} re-scaled each variable to unit variance with respect to its 3-hour temporal difference, averaged over space and time and noted that this had a large impact on overall model performance, a finding that was later corroborated by the authors of Stormer.
Furthermore, the loss function of global forecasts is commonly re-scaled per spatial location to account for differences in area per pixel induced by the sphere to grid mapping as in ERA5.
This spatial rescaling usually takes the form of the cosine of the latitude in models that utilise a rectangular grid, but is not necessary in e.g. CubedSphereNet and addressed by a geometry-dependent weighting in Spherical-FourCastNet.

GraphCast, CubedSphereNet and FourCastNet use the L2 loss, whereas PanGu-Weather and FuXi use the L1 loss.
Depending on the objective formulation of the models different temporal extents are used to compute the loss. 
PanGu trains each of their submodels for a different leadtime, but only includes a single step in the loss.
Most other models, such as CubedSphereNet, FourCastNet and GraphCast initially train on a single timestep, but towards the end of the training finetune on variable lengths of roll-outs, c.f. section \ref{sec:training}.
GraphCast further weights each variable by the standard deviation of the change between consecutive forecast steps to support uniform model performance across variables, they further note that additional weights on the per-variable losses can be used to encourage the model to focus on high impact variables like surface temperature. 
In the PanGu-Weather model a similar approach is taken to encourage uniform performance across variables, where the authors weight each variable inversely proportional to the average loss value in an early training run of the model.

\subsubsection{Probabilistic loss functions}
Probabilistic loss functions explicitly represent predictive uncertainty, allowing the model to predict fine-grained features and potential dynamics that are plausible, even if they differ from the observations.
Probabilistic forecasts are also desirable from an operational point of view \citep{slingo_uncertainty_2011} because the stakeholders that acquire these forecasts often use them to inform decision making processes that require risk assessment.
Therefore the calibration i.e., how well the predicted probabilities capture the true uncertainty, has to be investigated carefully \citep{graubner_calibration_2022}.

In order to leverage the advantages of probabilistic loss functions, the model needs to either provide mean and variance estimates or an ensemble of predicted states of all variables at all prediction times. In both cases, the model output is used to characterise its predictive distribution and a probabilistic loss is typically designed to quantify the likelihood of sampling the observed target from this distribution. One way of quantifying this is to use proper scoring rules, where the choice of scoring rule depends on the characteristics of the predictive distribution:
For distributions over continuous variables, it is common to use the Negative-Log-Likelihood (NLL) or the Continuous Ranked Probability Score (CRPS). 
In terms of implementation, it is convenient to assume that the predictive distribution is a normal distribution $\mathcal{N}(\mu, \sigma)$ with the parameters predicted by a DL model \citep{kendall_what_2017, stirn_variational_2020}. 
Alternatively, the statistics of an ensemble can be used directly to estimate the CRPS \citep{gneiting_strictly_2007}.
It should be noted that when facing input-dependent uncertainty variations, which naturally occur in weather dynamics, suitable normalisation techniques should be applied to NLL to avoid model collapse to highly uncertain predictions \citep{seitzer_pitfalls_2022}.
In the case of discretised variables, like bins of rainfall in MetNet, the Cross-Entropy loss has been used with great success.

Scoring rules like the CRPS are commonly used to verify ensemble forecasts and probabilistic forecasts in meteorology \citep{berrisch_crps_2021} and DLWP (e.g. in PanGu-Weather and CubedSphereNet).
However, only a few studies utilise them directly for training:
AtmoRep \citep{lessig_atmorep_2023} and Neural-GCM \citep{kochkov_neural_2023} report promising results using the statistics of ensemble predictions for training with first-order moment matching and CRPS, whereas FengWu and e.g. \cite{chapman_probabilistic_2022} output a mean and standard deviation directly for use with a NLL or CRPS loss, respectively.

\begin{figure}
    \centering
    \includegraphics[width=\textwidth]{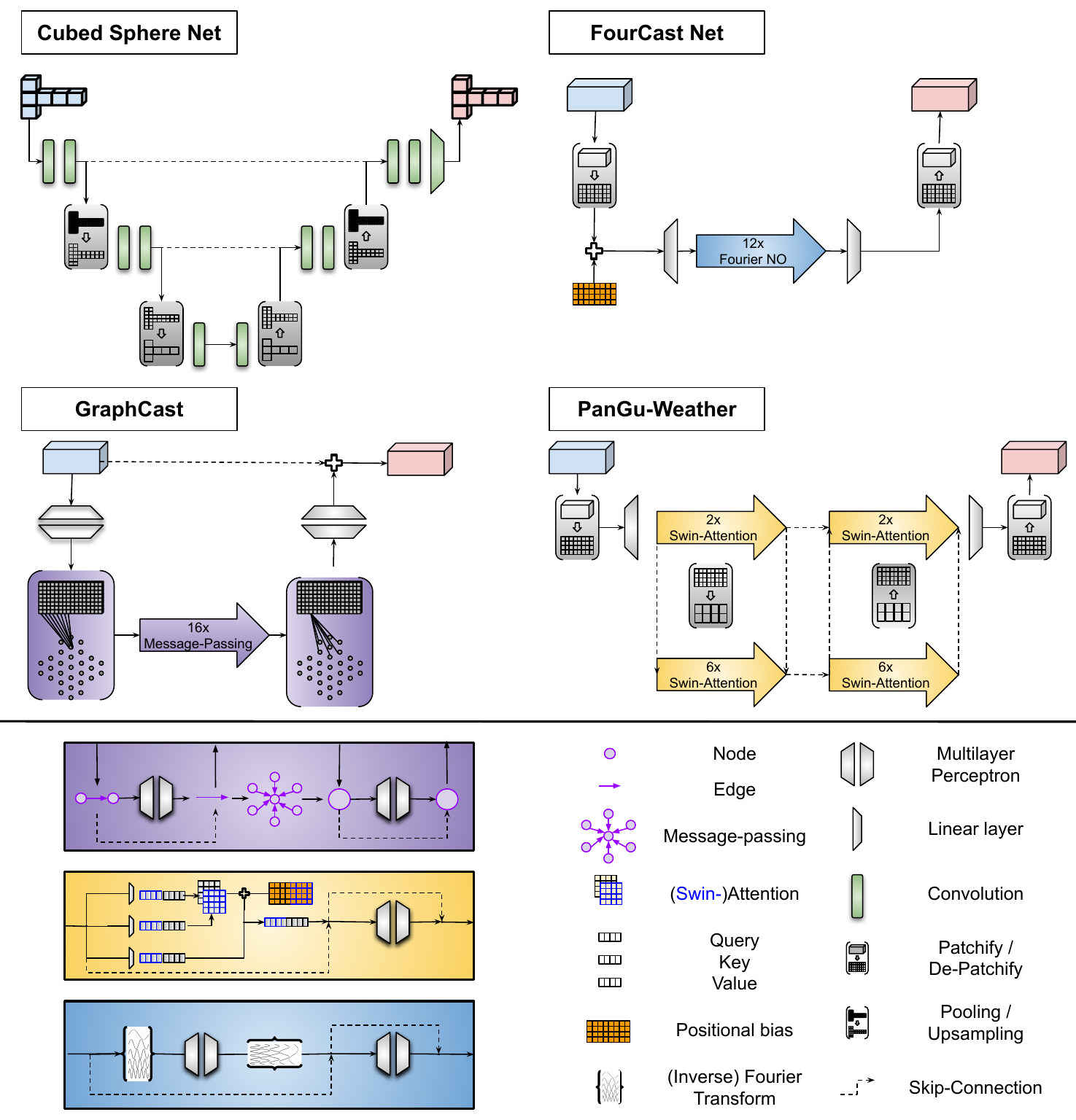}
    \caption{
    Architecture overview of the four reviewed models. 
    We emphasise encode-process-decode structures as well as the use of (multiple) latent spatial scales and/or representation formats. 
    Computational block designs are illustrated in the lower part of the figure, although we note that we omitted the ubiquitous use of LayerNorm in all of these blocks.
    For detailed explanations and illustrations of the architectures we refer to the respective figures in their original papers.
    }
    \label{fig:architectures}
\end{figure}

\subsection{Architecture design}
\label{sec:architectures}

\begin{ourbox}
\begin{custombox}{Neural Network Primitives}
\customlabel{box:nnp}{box 3}

Neural network models are designed to be highly modular, with a few computational blocks repeated across the entire architecture. 
Generally, each block consists of a linear map combined with a structural operation along the spatial or temporal data dimensions, followed by a non-linear activation function. 
The blocks usually also contain two components that vastly improve the training dynamics of deep models: a residual connection \citep{he_deep_2015, he_delving_2015} that adds the input of the block to its output, and a normalisation layer \citep{ioffe_batch_2015, ba_layer_2016, ulyanov_instance_2017, wu_group_2018} that shifts and scales the activations to be approximately standard normal.

\paragraph*{Convolution}
Convolutional Neural Networks (CNN \citep{lecun_convolutional_1995, lecun_gradient-based_1998}) operate by sliding a small window across an image, or any other regularly spaced grid, and computing the weighted sum of values inside of this window.
The well-known inductive bias of CNN is that they are translation equivariant \citep{bronstein_geometric_2021}, representing an assumption that the same processes occur everywhere and meaningful patterns are defined by relative differences.
Although CNNs are widely used and well-understood, they are limited by their receptive field---the area from which they can integrate information---requiring many consecutive layers to describe patterns across spatial scales.
Different approaches exist to alleviate this short-coming, for example the use of dilated filters \citep{yu_multi-scale_2016} or depthwise-separable convolutions \citep{chollet_xception_2017, liu_convnet_2022}.


\paragraph*{Message-Passing}
In Graph Neural Networks (GNN) spatiotemporal locations are taken to be the nodes of a graph and edges between these locations represent potential, pair-wise, interactions \citep{battaglia_relational_2018}.
By choosing the number of nodes, as well as their connectivity patterns, GNNs offer a more flexible structure for defining neighbourhoods on arbitrary geometric objects \citep{bronstein_geometric_2021}.
The message-passing framework \citep{gilmer_neural_2017} is used to model these interactions, by first computing a `message' that travels along each edge, and then integrating all incoming messages at a node.
Note that while GNN are often considered to be very data-efficient, i.e. they achieve high performance using few parameters and are therefore less prone to overfitting, they are computationally expensive and more challenging to utilise with standard deep learning libraries and GPU accelerators.


\paragraph*{Attention}
The Attention mechanism and the Transformer architecture in which it is embedded \citep{vaswani_attention_2017} has quickly established itself as the standard model for essentially all domains that are data-rich enough to support the high number of parameters required for them to function.
Attention mixes information across a set of tokens, by utilising the dot product similarity between all tokens to compute the weights of a weighted sum integrating information from all tokens at the same time \citep{tsai_transformer_2019}. 
This all-to-all mapping enables the processing of patterns on all scales simultaneously, as well as efficiently parallelised sequential predictions, however, it comes at a heavy computational cost, scaling quadratically with the number of tokens.



\paragraph*{Neural Operators}
The Fourier Neural Operator (FNO \cite{li_fourier_2022}) and its adaptive variation \citep{guibas_adaptive_2022} have caused a lot of excitement in the realm of deep learning for partial differential equations.
These models utilise the Discrete Fourier Transform to map the data from its spatial or temporal representation to the corresponding frequency domain, where a learned mapping can integrate information across all frequencies, and consequently, the entire input domain.
In contrast to Attention-based models, these models can achieve a global receptive field with quasi-linear computational complexity, while being invariant to the resolution of the underlying grid.


\end{custombox}
\end{ourbox}

Deep learning architectures are highly flexible and the ongoing exploration of their design space across all fields in which these architectures are applied has yielded a vast zoo of different model components and architectures (see box \ref{box:nnp} for a brief introduction to common model primitives).
In the following, we will break down this design space along the \textbf{Encode-Process-Decode} paradigm which informs the backbone structure of most modern architectures: \newline
Samples from the original data space are first  \textbf{encoded} in an abstract latent space, the resulting latent representations are \textbf{processed} by a deep neural network and finally the samples are \textbf{decoded} back to the data space.

\paragraph{Encoders}

The very first layer of any deep neural network fulfils one basic function of an encoder in that it projects from the data space into an abstract, almost always high-dimensional, latent space.
It is important to recognise that the utilisation of high-dimensional latent representations is a core tenet of deep learning (c.f. box \ref{box:dl}).
Conversely, the spatiotemporal resolution of the data will often be reduced in the initial encoding step, one can think of this as folding several measurements in space and time into the feature or variable dimension of the tensor that the network subsequently operates on.
Especially for weather fields it is crucial to compress the spatiotemporal extent of the data---since the internal activations of the network have to be stored for gradient backpropagation, the memory complexity scales with the product of each of the four spatiotemporal dimensions.

Among the reviewed models the most common form of initial encoding is the tokenisation scheme typically used in Vision Transformers \citep{dosovitskiy_image_2021}, where e.g. PanGu-Weather aggregates non-overlapping cubes of size (2x4x4) and (4x4) for upper-atmosphere or surface-level variables. 
This encoding reduces the spatial length of the data by a factor of 32 (respectively 16) and thereby enables the use of a computationally intense Transformer processor in the latent space.
The original FourCastNet, FuXi, FengWu as well as Stormer are built around the Vision Transformer architecture and use the same style of tokenisation.

CubedSphereNet and the Spherical-FourCastNet, on the other hand, do not utilise any initial compression, but instead rely on their computationally efficient processing units following a simple convolutional or linear encoder, respectively.

In graph-based neural networks such as GraphCast and KeislerNet the encoder functions by representing connections from multiple grid locations to the closest node on a coarse mesh \citep{sanchez-gonzalez_graph_2018} and aggregating them with a linear map, both compressing the data and highlighting another crucial utility of encoders: changing the format of the underlying data representation. 
GraphCast further uses an MLP encoder prior to the mesh-to-grid aggregation in order to unify the size of the latent representations of all node and edge features.

\paragraph{Processors}

Latent processors perform the bulk of neural computation, they iteratively refine and adapt the representations of the data such that the final decoder layer has the most salient information available to generate the desired prediction.
The representations of the data created by the intermediate latent processing steps are often considered highly abstract and do not, in general, correspond to physical quantities or allow for human interpretation.
We emphasise, however, that the characteristics of these representations are heavily shaped by the inductive biases of the objective, loss function and processing components used.

Most state-of-the-art DLWP architectures, including FourCastNet, GraphCast as well as PanGu-Weather and others, follow the design philosophy of high performance models in computer vision exemplified by the Vision Transformer \citep{dosovitskiy_image_2021} or the Swin-Transformer \citep{liu_swin_2021}.
In these models, a domain-specific encoding is followed by a large, homogeneous stack of processing blocks, without any further inductive biases incorporated past the initial encoder and subsequent choice of representational space.

CubedSphereNet and PanGu-Weather further include a hierarchy of multiple spatial scales within the processor, following the well-known design of the U-Net architecture \citep{ronneberger_u-net_2015}.
U-Net architectures consist of two network branches each with a cascade of down- or up-sampling operations and shortcuts that connect feature maps that share a spatial scale.
This multi-scale approach to architecture design has been prevalent in the DLWP literature for a long time and while the extent to which the features learned at different scales match dynamic processes on these scales, there are unanimous benefits to the computational efficiency of the model.
For a gentle introduction to the convolutional U-Net in a meteorological context we recommend \cite{ebert-uphoff_evaluation_2020} and \cite{lagerquist_using_2021}.
In a similar fashion, GraphCast uses multiple meshes, each with a different spatial resolution, which are processed jointly by the message-passing framework.

\paragraph{Decoders}

The final decoder stage of an architecture maps the abstract latent representation to the desired target space.
The nature of the target space is directly tied to the definition of the objective and it has been shown that an abstract latent space can often be decoded into several different output modalities such as time series, weather fields, or climate indices (see e.g. \cite{jaegle_perceiver_2022} which demonstrates this flexibility by combining videos and audio waveforms).
Since the goal of the decoder stage is \emph{not} to change the representation of the data, it is often times a simple linear map, although MLPs are used e.g. in GraphCast.

The decoder modules of the models we discuss here, differ in whether they decode the target field directly, or predict the \emph{difference} between the last input and the target frame.
In the context of video prediction it was hypothesised that it is more efficient to predict the transformation of the input, i.e. by applying a field of differences or a more complex vector field \citep{de_bezenac_deep_2018}.
This approach allows the model to maintain the current state information, potentially freeing up model capacity \citep{van_amersfoort_transformation-based_2017, luc_transformation-based_2020}.

The residual between two frames can also be interpreted as an estimate of the rate of change of the dynamics.
GraphCast predicts a single frame of residuals, whereas PanGu-Weather and FourCastNet produce a single, complete, target frame.
CubedSphereNet outputs two consecutive frames at once, implicitly encouraging the model to account for the rate of change between them.
A notable innovation put forward by the authors of PanGu-Weather is the use of separate decoders for upper air and surface variables, which could allow the individual decoders to be more tuned towards the specifics of the target variable.
The same approach was further extended to utilise separate decoders for all variables in FengWu.

\begin{figure}[!htb]
    \centering
    \includegraphics[width=\textwidth]{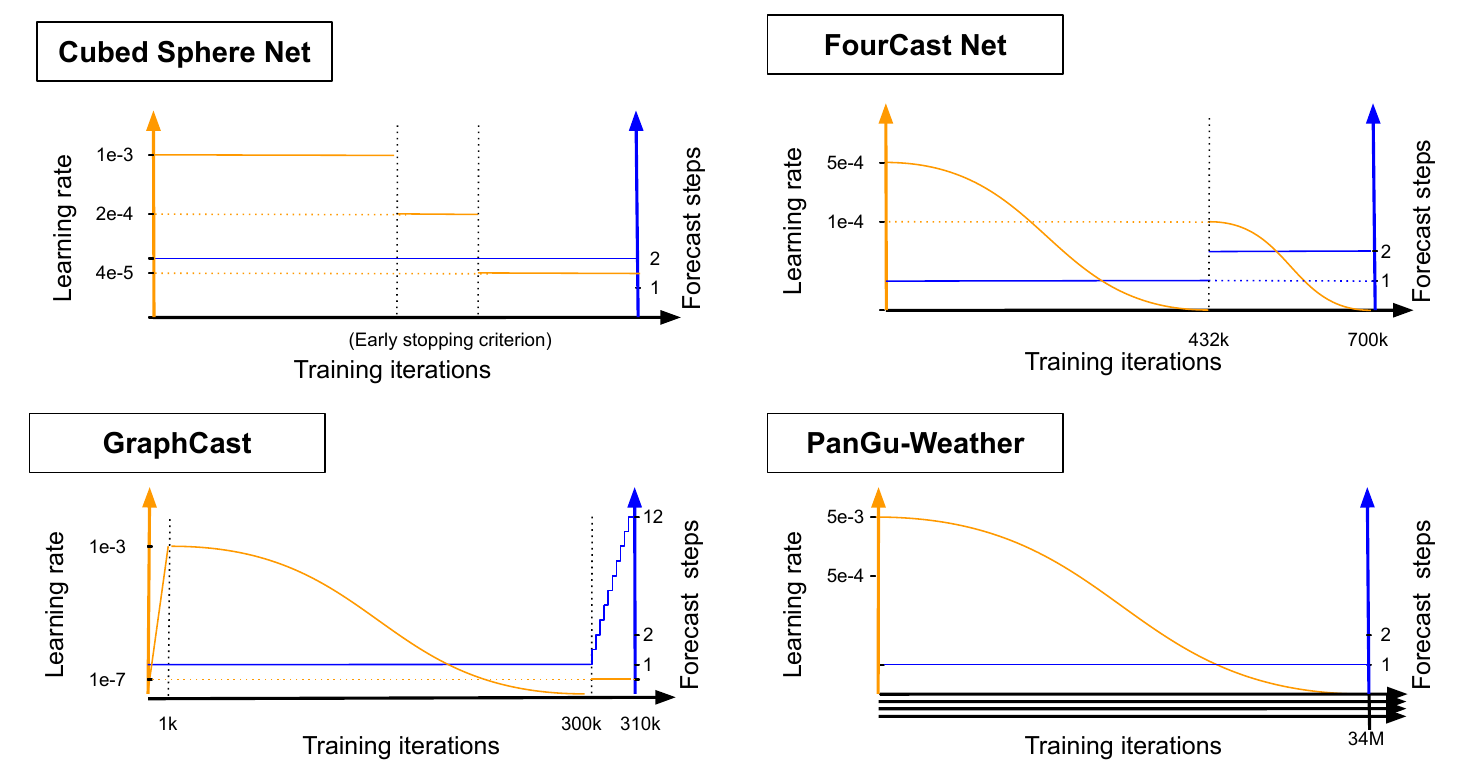}
    \caption{
    Training schemes of the four considered models. 
    GraphCast utilises a learning rate schedule consisting of a short linear warm-up, followed by a long cosine annealing phase and another short fine-tuning phase in the end. The number of autoregressive steps is initially kept at one but increased up to twelve in the final stage of training.
    PanGu-Weather is trained according only on one-step ahead predictions with a long cosine annealing schedule, we assume that the number of training iterations is different for each sub-model but note that this information has not been provided by the authors.
    FourCastNet uses a two-stage training procedure, first on one-step and then on two-step forecasts, with a cosine annealing schedule in each stage.
    CubedSphereNet does not use a pre-defined schedule, instead the learning rate is decreased by a factor of five once the validation criterion has not decreased for a given number of steps, we illustrate a possible trajectory here.
   }
    \label{fig:training}
\end{figure}

\subsection{Training}
\label{sec:training}

Once a suitable learning problem has been formalised, the parameters of the DL model are optimised such that they best represent the learning objective.
Neural networks are trained via stochastic gradient descent (SGD \cite{robbins_stochastic_1951}) on the surface of the loss function over the model parameters, where the optimisation landscape in each step depends on a small subset of the data which is called a (mini-)batch.
We briefly discuss intuitions about the loss surface, highlighting benefits of overparameterisation, and importance of learning rates---all of which are common to DLWP models.
Next, we discuss various curriculum learning strategies, defined as sequences of intermediate objectives that facilitate the learning of, e.g., complex spatiotemporal dynamics.

\subsubsection{Loss surfaces}

The overarching goal of optimisation is to find a minimum of the loss surface, i.e. a set of parameters for which the distance between the model output and the target is as small as possible.
In neural network architectures with many millions of parameters, all of which can have complex interdependencies, this surface can contain a multitude of local minima only a few of which will lead to good performance on unseen data. However, the mini-batch training with SGD has an inductive bias towards finding well-behaved solutions \citep{bartlett_deep_2021}.

Because the minimum of the training loss for one batch may be sub-optimal for another, SGD will only remain in flat regions of the loss surface that minimise a wide range of batches and which are thus more likely to generalise well to subsets of the data not seen during training \citep{hochreiter_flat_1997, keskar_large-batch_2017}.
A key feature of modern deep learning models that has been found to be highly synergistic with this style of optimisation is that they are heavily overparameterised and, therefore, it is often possible to find direct paths connecting the minima of several batches, which keeps the model from being stuck in suboptimal regions \citep{xing_walk_2018}.

In practice, the convergence of SGD towards a good optimum will depend strongly on the choice of hyperparameters, such as the learning rate and batch size, which jointly define the size of the update step \citep{goyal_accurate_2018}.
Additional methods have been introduced on top of SGD to either accelerate the traversal of the loss surface or improve generalisation through regularisation. 
Noteworthy are the use of momentum terms \citep{sutskever_importance_2013} and adaptive step sizes \citep{duchi_adaptive_2011} which are combined in the popular Adam optimisation algorithm \citep{kingma_adam_2017}.
AdamW \citep{loshchilov_decoupled_2019} further augments Adam by incorporating the well-known weight decay regularisation technique \citep{hanson_comparing_nodate}, restricting the magnitude of network parameters to a ball around the origin of the parameter space.

Besides weight decay, another well-established regularisation method are various forms of dropout \citep{srivastava_dropout_2014} that operate by stochastically removing a fraction of parameters from the forward computation, preventing them from co-adapting.
Through the lens of flat minima discussed earlier, we can also understand dropout as creating a complementary form of subsampling of the parameter-loss-surface.
Other forms of dropout incorporate additional structure, for example by dropping out entire computational blocks \citep{ghiasi_dropblock_2018, wu_blockdrop_2019}, or structured fractions of the input \citep{he_masked_2021}.

Adam and AdamW are the prevalent optimisers, being used in all of the reviewed DLWP models. 
Batch-sizes are generally chosen to best utilise the available compute resources with learning rates often defaulting to values around the order of $1 \times 10^{-4}$.
GraphCast and PanGu-Weather include weight decay terms and PanGu-Weather utilises DropPath as additional regularisation method.
The number of parameters used in current DLWP models covers a wide range, with CubedSphereNet being on the order a few million, GraphCast on the order of tens of millions and FourCastNet and PanGu-Weather having hundreds of millions of parameters.
AtmoRep recently became the first DLWP model that exceeded 1 billion parameters \citep{lessig_atmorep_2023}.


\subsubsection{Curriculum learning}

Often times it is beneficial to alter the objective throughout the optimisation procedure \citep{bengio_curriculum_2009}---for example by initially training on less compute-intensive or easier to solve tasks---an idea that is consistent with how human learning works.
Beginning training with an easier-to-solve task may also be seen as a sophisticated initialisation of model weights, beginning optimisation from a well-performing, instead of random, model configuration \citep{ erhan_difficulty_2009}.
This pre-training approach is prevalent throughout modern DL and has led to the recent emergence of foundational models that can be easily `fine-tuned' to many different tasks e.g. the Generative Pre-trained Transformer (GPT \cite{brown_language_2020}).

When designing a curriculum it is important to consider that neural networks can easily forget tasks they previously learned \citep{mccloskey_catastrophic_1989}.
A good curriculum strategy will therefore utilise the previously learned features to more efficiently solve the following task---when modelling high resolution spatiotemporal data, curriculum learning strategies may optimise only one-step ahead predictions, or operate on coarse spatial resolutions, initially, before including more expensive and complex dynamics once the model has achieved a base level of competence.
Similarly, the prediction of challenging diagnostic variables such as precipitation many days in advance, may be more suitable to be learned once the large-scale features of the atmospheric state are adequately propagated forward in time.

The success of a curriculum learning approach strongly depends on the choice of learning rate---initially, a large learning rate can favour exploration of the parameter space and accelerate the model towards suitable solutions \citep{smith_super-convergence_2018}, whereas in the later fine-tuning stages the model should stay within a suitable region and converge to the best local minimum.
Often times the learning rate is therefore scheduled to increase or decrease at pre-determined intervals, e.g. following the cosine annealing scheme proposed in \citep{loshchilov_sgdr_2017} or the cyclical schemes proposed by \citep{smith_cyclical_2017}.

Both GraphCast and FourCastNet utilise curriculum learning strategies where they initially train the model only on one-step ahead predictions and only in a second phase of training do auto-regressive predictions using the improved model outputs as subsequent inputs.
GraphCast further fine-tunes its long-term predictions in a brief final training stage with up to twelve auto-regressive steps.
FourCastNet on the other hand adds a small network to predict precipitation using the outputs of the base model, but only trains this sub-network in a final fine-tuning stage.
In either paper, the learning rate is scheduled, with GraphCast initially increasing, annealing and then fine-tuning with a close-to-zero learning rate, whereas FourCastNet anneals its learning rate within each stage and then restarts training with a higher learning rate (c.f. figure \ref{fig:training}).
Learning rate annealing in both models, as well as PanGu-Weather, is done using a cosine annealing schedule.

\section{Discussion}

The rise of deep learning in weather prediction over the last few years has been quick and unexpected---a common perspective by experts in the field was that the medium range of weather forecasting was too challenging and the existing numerical models too strong to compete against \citep{chantry_opportunities_2021}.
As it turns out, the strengths of numerical models can be utilised by DLWP, building on the detailed training data and high-powered compute infrastructure already in place.
At the time of writing, several DLWP models, namely PanGu-Weather, GraphCast, FourCastNet as well as FuXi are run operationally by the European Center for Medium Range Weather Forecasting (ECMWF)---highlighting the ease and efficiency of running DL models at inference time---and to date their forecasts have been of similar quality to state-of-the-art NWP forecasts with some noticeable spectral- and intensity biases, especially at longer lead-times.
Operationalisation will further clarify merits and limitations of these models in the years to come, while the rapid pace of development in deep learning will likely improve the capabilities of DLWP.
For the purpose of this review, we are interested in the most salient design choices common to current models and a perspective where future methodological development may lead us---in the following we will discuss both aspects in turn.

\subsection{Where we are right now}

When analysing the design choices made in the current generation of DLWP models, we notice compute efficiency as the clear guiding principle. Domain-specific inductive biases play only minor roles, indicating that the high quality of training data available is sufficient to avoid defective model behaviour.

Atmospheric dynamics are well-known to consist of salient features across all scales in three-dimensional space as well as time.
Indeed, models that were trained on data with higher resolutions and larger state vectors tend to produce more accurate forecasts, with each additional field adding both new information as well as physical constraints on the unfolding dynamics.

Objective design (figure~\ref{fig:objectives}) highlights the need to trade-off between increased resolution and managing compute requirements further---high resolution models are largely trained to deterministically predict only one or two steps forward in time, a choice that closely resembles traditional NWP.
Deep learning offers substantial flexibility to capture more flexible time-stepping schemes, as well as probabilistic and generative models, which could further enhance and complement existing approaches to weather prediction.

Pixel-wise L1- and L2-norms are the standard loss function (figure~\ref{fig:losses}) to use, but their issues have long been recognised \citep{gilleland_intercomparison_2009} and are a likely cause for the observed deficiencies of DLWP models in terms of long-term spectral power and low-intensity biases for extreme events.
Likelihood-based losses are better suited to capture uncertainty and rare events, but are nevertheless constrained by the assumption that individual pixels are statistically independent of each other.
Computing the loss in the feature-space of a secondary neural network \citep{zhang_deep_2017, mathieu_deep_2016} has been shown to largely alleviate these issues, also in the weather and climate domain \citep{hoffmann_atmodist_2022, hess_deep_2022}.

Neural network design (figure~\ref{fig:architectures}) follows typical architecture schemes, with shallow encoders and decoders, homogeneous stacks of computational blocks as processors and, occasionally, U-Net structures.
We note that most models we reviewed utilise a strong spatial (de)compression in their encoder (decoder)---often operating on effective internal resolutions that are quite low compared to the data resolution---and that these encoders contain most of the domain-specific choices, i.e., which fields and vertical levels to treat jointly or how to represent the spherical geometry of the earth.
\newline
Optimisation methods (figure~\ref{fig:training}) are fairly homogeneous with AdamW and Cosine-Annealing being prevalent with hyper-parameters largely following industry standards.
The choice of training curriculum appears to be explored more widely, with the main unifying component being a pretraining stage that only considers next-step prediction without any auto-regressive component.
Computationally expensive generative components, or multi-step objectives, are only added towards the end of training once the model backbone performs at an appreciable level.

\subsection{The road ahead}

Across domains deep learning models have been found to capture even the most complex physical interactions, at least when provided with sufficient observations and compute resources to train the largest possible models.
Although the existence of neural scaling laws \citep{kaplan_scaling_2020} is, at the moment, rather an empirical observation than a theoretical law, these observations suggest a power law relation between the available data, compute and model complexity.
In the weather and climate domain, neither of these dimensions has been fully exhausted yet and thus we expect further improvements in the skill of DLWP, from short- to medium- or even climatic timescales.
Throughout this review we were concerned with medium-range forecasting models, which were trained on the ERA5 dataset consisting of reanalysis based on the NWP model that the DLWP models subsequently outperformed. What enables the DLWP models to outperform the NWP model underlying their training data?

We suspect that the learned \emph{representations} of any individual atmospheric state are enriched by the statistical memory of the neural network, thus utilising information beyond what is contained in the initial state.
Key to this hypothesis is the fact that the estimate of an atmospheric state used for forecasting, by necessity, is incomplete, processes occurring below the discretisation grid---such as convection or land-atmosphere interactions, as well as small-scale atmospheric turbulence ---can not be fully captured and necessitate a statistical perspective \citep{palmer_primacy_2017}.
By flexibly integrating estimates of these sub-grid effects, as well as analogues of atmospheric trajectories, into their forward process, DLWP models could feasibly outperform the numerical teachers they emulate.

From this perspective, we are intrigued by two complementary lines of research: improving the forward process of the models for more stable and physically plausible emulators and building on the statistical estimation capacities towards foundational generative models.

Physics-informed deep learning \citep{kashinath_physics-informed_2021} and various forms of neural solvers for PDEs \citep{cuomo_2022} have emerged in the last few years and are an active area of research.
Indeed, many DLWP models are formulated closely in spirit to NWP models, although the degree to which physical equations are represented in these models varies, mainly drawing on deep learning to improve the efficiency of the numerical models.
For example, GraphCast and other Message-Passing networks can be seen as a learned Finite Element Method \citep{alet_graph_2019} although they do not include any explicit physical knowledge.
Other approaches, such as ACE \citep{watt-meyer_ace_2023}, utilise (Spherical)-FNO to generate long-term stable roll-outs of atmospheric dynamics, again, without an explicit representation of the underlying PDEs, whereas the recently published Neural-GCM \citep{kochkov_neural_2023} explicitly integrates a dynamical core with a spectral solver and a neural network trained jointly to predict the parameterisation.

Generative models, on the other hand, come from the perspective of Bayesian modelling, aiming to capture the underlying data distribution as closely as possible. They thus describe patterns of correlation across space and time, rather than an explicit dynamical process.
Many of the current successes of deep learning in language and vision are attributed to generatively, or self-supervised, pre-trained models.
Rather than producing the most effective solution to any individual task, these approaches excel as a foundation for further fine-tuning and specialisation.
AtmoRep \citep{lessig_atmorep_2023} highlights the potential of such models in earth system modelling to be used for a variety of tasks, including forecasting, bias correction, statistical downscaling or the integration of different data modalities.

Forecasting with generative models has also emerged as the leading paradigm for video prediction \citep{denton_stochastic_2018, liu_sora_2024}, including VAE or Diffusion based components in recurrent forward models.
The generative component is often utilised to inject noise into the forward process, creating a variety of realistic trajectories given an initial state---a task that bears striking similarity to the approach of stochastically perturbed parameterisations in NWP \citep{buizza_stochastic_1999}.
Current front-runners in terms of skillful long-term prediction, rather than stable long-term simulations as the approaches discussed above, follow the same paradigm.
Examples are the use of Diffusion-based perturbation in GenCast \citep{price_gencast_2023}, the VAE-like noise in FuXi-S2S \citep{chen_fuxi-s2s_2023}, and the noise-injection of Neural-GCM \citep{kochkov_neural_2023}.


\subsection{Concluding remarks}
We reviewed currently impactful approaches that showcase the wide range of applicable methodology from the broader DL community and how these methods integrate weather-specific knowledge into their design.
The presented template for discussing design choices is widely applicable beyond the models reviewed here.
Although promising, the current generation of models are likely to be superseded by more accurate, efficient, and probabilistic forecasting models in the near future.
Especially in the face of anthropogenic climate change and the resulting increase in the likelihood and intensity of extreme weather events, it is critical that weather forecasting models are as skilful as they can be to decrease the consequent humanitarian, biological, and economic losses.


\section*{Acknowledgements}
The authors acknowledge funding from the intramural project Modeling and Understanding SpatioTemporal Environmental INteractions (MUSTEIN) via the Machine Learning Cluster of Excellence funded by the German Research Foundation (DFG) under the German Excellence Strategy EXC 2064/1 - 390727645. 
It received additional support by a fellowship within the IFI programme of the German Academic Exchange Service (DAAD). 
We acknowledge the International Max Planck Research School for Intelligent Systems (IMPRS-IS) for supporting Matthias Karlbauer. 
Martin Butz and Sebastian Otte acknowledge additional funding from the Cyber Valley in Tübingen, CyVy-RF-2020-15.
The authors would also like to thank Manuel Traub, Fedor Scholz, Florian Ebmeier, Tobias Menge, Jakob Schlör, Felix Strnad, Moritz Haas, and Gwen Hirsch for invaluable discussions and feedback on the manuscript.

\bibliography{references}

\begin{thebibliography}{134}
\providecommand{\natexlab}[1]{#1}
\providecommand{\url}[1]{\texttt{#1}}
\expandafter\ifx\csname urlstyle\endcsname\relax
  \providecommand{\doi}[1]{doi: #1}\else
  \providecommand{\doi}{doi: \begingroup \urlstyle{rm}\Url}\fi

\bibitem[Stott(2016)]{stott_how_2016}
Peter Stott.
\newblock How climate change affects extreme weather events.
\newblock \emph{Science}, 352\penalty0 (6293):\penalty0 1517--1518, June 2016.
\newblock \doi{10.1126/science.aaf7271}.

\bibitem[Clarke et~al.(2022)Clarke, Otto, Stuart-Smith, and
  Harrington]{clarke_extreme_2022}
Ben Clarke, Friederike Otto, Rupert Stuart-Smith, and Luke Harrington.
\newblock Extreme weather impacts of climate change: an attribution
  perspective.
\newblock \emph{Environmental Research: Climate}, 1\penalty0 (1):\penalty0
  012001, June 2022.
\newblock ISSN 2752-5295.
\newblock \doi{10.1088/2752-5295/ac6e7d}.
\newblock URL \url{https://dx.doi.org/10.1088/2752-5295/ac6e7d}.
\newblock Publisher: IOP Publishing.

\bibitem[Bauer et~al.(2015)Bauer, Thorpe, and Brunet]{bauer_quiet_2015}
Peter Bauer, Alan Thorpe, and Gilbert Brunet.
\newblock The quiet revolution of numerical weather prediction.
\newblock \emph{Nature}, 525\penalty0 (7567):\penalty0 47--55, September 2015.
\newblock ISSN 1476-4687.
\newblock \doi{10.1038/nature14956}.
\newblock URL \url{https://www.nature.com/articles/nature14956}.
\newblock Bandiera\_abtest: a Cg\_type: Nature Research Journals Number: 7567
  Primary\_atype: Reviews Publisher: Nature Publishing Group Subject\_term:
  Atmospheric dynamics;Climate sciences Subject\_term\_id:
  atmospheric-dynamics;climate-sciences.

\bibitem[Bauer et~al.(2021)Bauer, Dueben, Hoefler, Quintino, Schulthess, and
  Wedi]{bauer_digital_2021}
Peter Bauer, Peter~D. Dueben, Torsten Hoefler, Tiago Quintino, Thomas~C.
  Schulthess, and Nils~P. Wedi.
\newblock The digital revolution of {Earth}-system science.
\newblock \emph{Nature Computational Science}, 1\penalty0 (2):\penalty0
  104--113, February 2021.
\newblock ISSN 2662-8457.
\newblock \doi{10.1038/s43588-021-00023-0}.
\newblock URL \url{http://www.nature.com/articles/s43588-021-00023-0}.

\bibitem[Kalnay(2002)]{kalnay_atmospheric_2002}
Eugenia Kalnay.
\newblock Atmospheric {Modeling}, {Data} {Assimilation} and {Predictability},
  November 2002.
\newblock URL
  \url{https://www.cambridge.org/highereducation/books/atmospheric-modeling-data-assimilation-and-predictability/C5FD207439132836E85027754CE9BC1A}.
\newblock ISBN: 9780511802270 Publisher: Cambridge University Press.

\bibitem[Stensrud(2007)]{stensrud_parameterization_2007}
David~J. Stensrud.
\newblock \emph{Parameterization {Schemes}: {Keys} to {Understanding}
  {Numerical} {Weather} {Prediction} {Models}}.
\newblock Cambridge University Press, Cambridge, 2007.
\newblock ISBN 978-0-521-12676-2.
\newblock \doi{10.1017/CBO9780511812590}.
\newblock URL
  \url{https://www.cambridge.org/core/books/parameterization-schemes/C7C8EC8901957314433BE7C8BC36F16D}.

\bibitem[Schneider et~al.(2017)Schneider, Lan, Stuart, and
  Teixeira]{schneider_earth_2017}
Tapio Schneider, Shiwei Lan, Andrew Stuart, and João Teixeira.
\newblock Earth {System} {Modeling} 2.0: {A} {Blueprint} for {Models} {That}
  {Learn} {From} {Observations} and {Targeted} {High}-{Resolution}
  {Simulations}.
\newblock \emph{Geophysical Research Letters}, 44\penalty0 (24), December 2017.
\newblock ISSN 0094-8276, 1944-8007.
\newblock \doi{10.1002/2017GL076101}.
\newblock URL \url{http://arxiv.org/abs/1709.00037}.
\newblock arXiv: 1709.00037.

\bibitem[Slingo and Palmer(2011)]{slingo_uncertainty_2011}
Julia Slingo and Tim Palmer.
\newblock Uncertainty in weather and climate prediction.
\newblock \emph{Philosophical Transactions of the Royal Society A:
  Mathematical, Physical and Engineering Sciences}, 369\penalty0
  (1956):\penalty0 4751--4767, December 2011.
\newblock \doi{10.1098/rsta.2011.0161}.
\newblock URL
  \url{https://royalsocietypublishing.org/doi/10.1098/rsta.2011.0161}.
\newblock Publisher: Royal Society.

\bibitem[Bannister(2017)]{bannister_review_2017}
R.~N. Bannister.
\newblock A review of operational methods of variational and
  ensemble-variational data assimilation.
\newblock \emph{Quarterly Journal of the Royal Meteorological Society},
  143\penalty0 (703):\penalty0 607--633, 2017.
\newblock ISSN 1477-870X.
\newblock \doi{10.1002/qj.2982}.
\newblock URL \url{https://onlinelibrary.wiley.com/doi/abs/10.1002/qj.2982}.

\bibitem[Ben-Bouallegue et~al.(2023)Ben-Bouallegue, Clare, Magnusson, Gascon,
  Maier-Gerber, Janousek, Rodwell, Pinault, Dramsch, Lang, Raoult, Rabier,
  Chevallier, Sandu, Dueben, Chantry, and
  Pappenberger]{ben-bouallegue_rise_2023}
Zied Ben-Bouallegue, Mariana C.~A. Clare, Linus Magnusson, Estibaliz Gascon,
  Michael Maier-Gerber, Martin Janousek, Mark Rodwell, Florian Pinault,
  Jesper~S. Dramsch, Simon T.~K. Lang, Baudouin Raoult, Florence Rabier,
  Matthieu Chevallier, Irina Sandu, Peter Dueben, Matthew Chantry, and Florian
  Pappenberger.
\newblock The rise of data-driven weather forecasting, November 2023.
\newblock URL \url{http://arxiv.org/abs/2307.10128}.
\newblock arXiv:2307.10128 [physics].

\bibitem[Shi et~al.(2015)Shi, Chen, Wang, Yeung, Wong, and
  Woo]{shi_convolutional_2015}
Xingjian Shi, Zhourong Chen, Hao Wang, Dit-Yan Yeung, Wai-kin Wong, and
  Wang-chun Woo.
\newblock Convolutional lstm network: A machine learning approach for
  precipitation nowcasting.
\newblock In \emph{Proceedings of the 28th International Conference on Neural
  Information Processing Systems - Volume 1}, NIPS'15, page 802–810,
  Cambridge, MA, USA, 2015. MIT Press.

\bibitem[Dueben and Bauer(2018)]{dueben_challenges_2018}
Peter~D. Dueben and Peter Bauer.
\newblock Challenges and design choices for global weather and climate models
  based on machine learning.
\newblock \emph{Geoscientific Model Development}, 11\penalty0 (10):\penalty0
  3999--4009, October 2018.
\newblock ISSN 1991-9603.
\newblock \doi{10.5194/gmd-11-3999-2018}.
\newblock URL \url{https://gmd.copernicus.org/articles/11/3999/2018/}.

\bibitem[Weyn et~al.(2019)Weyn, Durran, and Caruana]{weyn_can_2019}
Jonathan~A. Weyn, Dale~R. Durran, and Rich Caruana.
\newblock Can {Machines} {Learn} to {Predict} {Weather}? {Using} {Deep}
  {Learning} to {Predict} {Gridded} 500-{hPa} {Geopotential} {Height} {From}
  {Historical} {Weather} {Data}.
\newblock \emph{Journal of Advances in Modeling Earth Systems}, 11\penalty0
  (8):\penalty0 2680--2693, 2019.
\newblock ISSN 1942-2466.
\newblock \doi{10.1029/2019MS001705}.
\newblock URL
  \url{https://onlinelibrary.wiley.com/doi/abs/10.1029/2019MS001705}.

\bibitem[Scher(2018)]{scher_toward_2018}
S.~Scher.
\newblock Toward {Data}-{Driven} {Weather} and {Climate} {Forecasting}:
  {Approximating} a {Simple} {General} {Circulation} {Model} {With} {Deep}
  {Learning}.
\newblock \emph{Geophysical Research Letters}, 45\penalty0 (22):\penalty0
  12,616--12,622, 2018.
\newblock ISSN 1944-8007.
\newblock \doi{10.1029/2018GL080704}.
\newblock URL
  \url{https://onlinelibrary.wiley.com/doi/abs/10.1029/2018GL080704}.

\bibitem[Kurth et~al.(2022)Kurth, Subramanian, Harrington, Pathak, Mardani,
  Hall, Miele, Kashinath, and Anandkumar]{kurth_fourcastnet_2022}
Thorsten Kurth, Shashank Subramanian, Peter Harrington, Jaideep Pathak, Morteza
  Mardani, David Hall, Andrea Miele, Karthik Kashinath, and Animashree
  Anandkumar.
\newblock {FourCastNet}: {Accelerating} {Global} {High}-{Resolution} {Weather}
  {Forecasting} using {Adaptive} {Fourier} {Neural} {Operators}, August 2022.
\newblock URL \url{http://arxiv.org/abs/2208.05419}.
\newblock arXiv:2208.05419 [physics].

\bibitem[LeCun et~al.(2015)LeCun, Bengio, and Hinton]{lecun_deep_2015}
Yann LeCun, Yoshua Bengio, and Geoffrey Hinton.
\newblock Deep learning.
\newblock \emph{Nature}, 521\penalty0 (7553):\penalty0 436--444, May 2015.
\newblock ISSN 1476-4687.
\newblock \doi{10.1038/nature14539}.
\newblock URL \url{https://www.nature.com/articles/nature14539}.
\newblock Number: 7553 Publisher: Nature Publishing Group.

\bibitem[Bengio et~al.(2013)Bengio, Courville, and
  Vincent]{bengio_representation_2014}
Yoshua Bengio, Aaron Courville, and Pascal Vincent.
\newblock Representation learning: a review and new perspectives.
\newblock \emph{IEEE transactions on pattern analysis and machine
  intelligence}, 35\penalty0 (8):\penalty0 1798--1828, August 2013.
\newblock ISSN 0162-8828.
\newblock \doi{10.1109/tpami.2013.50}.

\bibitem[Wolpert and Macready(1997)]{wolpert_no_1997}
D.H. Wolpert and W.G. Macready.
\newblock No free lunch theorems for optimization.
\newblock \emph{IEEE Transactions on Evolutionary Computation}, 1\penalty0
  (1):\penalty0 67--82, April 1997.
\newblock ISSN 1941-0026.
\newblock \doi{10.1109/4235.585893}.
\newblock Conference Name: IEEE Transactions on Evolutionary Computation.

\bibitem[Battaglia et~al.(2018)Battaglia, Hamrick, Bapst, Sanchez-Gonzalez,
  Zambaldi, Malinowski, Tacchetti, Raposo, Santoro, Faulkner, Gulcehre, Song,
  Ballard, Gilmer, Dahl, Vaswani, Allen, Nash, Langston, Dyer, Heess, Wierstra,
  Kohli, Botvinick, Vinyals, Li, and Pascanu]{battaglia_relational_2018}
Peter~W. Battaglia, Jessica~B. Hamrick, Victor Bapst, Alvaro Sanchez-Gonzalez,
  Vinicius Zambaldi, Mateusz Malinowski, Andrea Tacchetti, David Raposo, Adam
  Santoro, Ryan Faulkner, Caglar Gulcehre, Francis Song, Andrew Ballard, Justin
  Gilmer, George Dahl, Ashish Vaswani, Kelsey Allen, Charles Nash, Victoria
  Langston, Chris Dyer, Nicolas Heess, Daan Wierstra, Pushmeet Kohli, Matt
  Botvinick, Oriol Vinyals, Yujia Li, and Razvan Pascanu.
\newblock Relational inductive biases, deep learning, and graph networks.
\newblock \emph{arXiv:1806.01261 [cs, stat]}, October 2018.
\newblock URL \url{http://arxiv.org/abs/1806.01261}.
\newblock arXiv: 1806.01261.

\bibitem[Weyn et~al.(2021)Weyn, Durran, Caruana, and
  Cresswell-Clay]{weyn_sub-seasonal_2021}
Jonathan~A. Weyn, Dale~R. Durran, Rich Caruana, and Nathaniel Cresswell-Clay.
\newblock Sub-{Seasonal} {Forecasting} {With} a {Large} {Ensemble} of
  {Deep}-{Learning} {Weather} {Prediction} {Models}.
\newblock \emph{Journal of Advances in Modeling Earth Systems}, 13\penalty0
  (7):\penalty0 e2021MS002502, 2021.
\newblock ISSN 1942-2466.
\newblock \doi{10.1029/2021MS002502}.
\newblock URL
  \url{https://agupubs.onlinelibrary.wiley.com/doi/abs/10.1029/2021MS002502}.

\bibitem[Karlbauer et~al.(2023)Karlbauer, Cresswell-Clay, Moreno, Durran,
  Kurth, and Butz]{karlbauer2023advancing}
Matthias Karlbauer, Nathaniel Cresswell-Clay, Raul~A. Moreno, Dale~R. Durran,
  Thorsten Kurth, and Martin~V. Butz.
\newblock Advancing parsimonious deep learning weather prediction using the
  healpix mesh, 2023.

\bibitem[Weyn et~al.(2020)Weyn, Durran, and Caruana]{weyn_improving_2020}
Jonathan~A. Weyn, Dale~R. Durran, and Rich Caruana.
\newblock Improving {Data}-{Driven} {Global} {Weather} {Prediction} {Using}
  {Deep} {Convolutional} {Neural} {Networks} on a {Cubed} {Sphere}.
\newblock \emph{Journal of Advances in Modeling Earth Systems}, 12\penalty0
  (9):\penalty0 e2020MS002109, 2020.
\newblock ISSN 1942-2466.
\newblock \doi{10.1029/2020MS002109}.
\newblock URL
  \url{https://onlinelibrary.wiley.com/doi/abs/10.1029/2020MS002109}.

\bibitem[Pathak et~al.(2022)Pathak, Subramanian, Harrington, Raja,
  Chattopadhyay, Mardani, Kurth, Hall, Li, Azizzadenesheli, Hassanzadeh,
  Kashinath, and Anandkumar]{pathak_fourcastnet_2022}
Jaideep Pathak, Shashank Subramanian, Peter Harrington, Sanjeev Raja, Ashesh
  Chattopadhyay, Morteza Mardani, Thorsten Kurth, David Hall, Zongyi Li, Kamyar
  Azizzadenesheli, Pedram Hassanzadeh, Karthik Kashinath, and Animashree
  Anandkumar.
\newblock {FourCastNet}: {A} {Global} {Data}-driven {High}-resolution {Weather}
  {Model} using {Adaptive} {Fourier} {Neural} {Operators}, February 2022.
\newblock URL \url{http://arxiv.org/abs/2202.11214}.
\newblock Number: arXiv:2202.11214 arXiv:2202.11214 [physics].

\bibitem[Bonev et~al.(2023)Bonev, Kurth, Hundt, Pathak, Baust, Kashinath, and
  Anandkumar]{bonev_spherical_2023}
Boris Bonev, Thorsten Kurth, Christian Hundt, Jaideep Pathak, Maximilian Baust,
  Karthik Kashinath, and Anima Anandkumar.
\newblock Spherical {Fourier} {Neural} {Operators}: {Learning} {Stable}
  {Dynamics} on the {Sphere}, June 2023.
\newblock URL \url{http://arxiv.org/abs/2306.03838}.
\newblock arXiv:2306.03838 [physics].

\bibitem[Keisler(2022)]{keisler_forecasting_2022}
Ryan Keisler.
\newblock Forecasting {Global} {Weather} with {Graph} {Neural} {Networks},
  February 2022.
\newblock URL \url{http://arxiv.org/abs/2202.07575}.
\newblock Number: arXiv:2202.07575 arXiv:2202.07575 [physics].

\bibitem[Lam et~al.(2022)Lam, Sanchez-Gonzalez, Willson, Wirnsberger,
  Fortunato, Pritzel, Ravuri, Ewalds, Alet, Eaton-Rosen, Hu, Merose, Hoyer,
  Holland, Stott, Vinyals, Mohamed, and Battaglia]{lam_graphcast_2022}
Remi Lam, Alvaro Sanchez-Gonzalez, Matthew Willson, Peter Wirnsberger, Meire
  Fortunato, Alexander Pritzel, Suman Ravuri, Timo Ewalds, Ferran Alet, Zach
  Eaton-Rosen, Weihua Hu, Alexander Merose, Stephan Hoyer, George Holland,
  Jacklynn Stott, Oriol Vinyals, Shakir Mohamed, and Peter Battaglia.
\newblock {GraphCast}: {Learning} skillful medium-range global weather
  forecasting, December 2022.
\newblock URL \url{http://arxiv.org/abs/2212.12794}.
\newblock arXiv:2212.12794 [physics].

\bibitem[Bi et~al.(2022)Bi, Xie, Zhang, Chen, Gu, and
  Tian]{bi_pangu-weather_2022}
Kaifeng Bi, Lingxi Xie, Hengheng Zhang, Xin Chen, Xiaotao Gu, and Qi~Tian.
\newblock Pangu-{Weather}: {A} {3D} {High}-{Resolution} {Model} for {Fast} and
  {Accurate} {Global} {Weather} {Forecast}, November 2022.
\newblock URL \url{http://arxiv.org/abs/2211.02556}.
\newblock arXiv:2211.02556 [physics].

\bibitem[Chen et~al.(2023{\natexlab{a}})Chen, Zhong, Zhang, Cheng, Xu, Qi, and
  Li]{chen_fuxi_2023}
Lei Chen, Xiaohui Zhong, Feng Zhang, Yuan Cheng, Yinghui Xu, Yuan Qi, and Hao
  Li.
\newblock {FuXi}: {A} cascade machine learning forecasting system for 15-day
  global weather forecast, October 2023{\natexlab{a}}.
\newblock URL \url{http://arxiv.org/abs/2306.12873}.
\newblock arXiv:2306.12873 [physics].

\bibitem[Chen et~al.(2023{\natexlab{b}})Chen, Han, Gong, Bai, Ling, Luo, Chen,
  Ma, Zhang, Su, Ci, Li, Yang, and Ouyang]{chen_fengwu_2023}
Kang Chen, Tao Han, Junchao Gong, Lei Bai, Fenghua Ling, Jing-Jia Luo, Xi~Chen,
  Leiming Ma, Tianning Zhang, Rui Su, Yuanzheng Ci, Bin Li, Xiaokang Yang, and
  Wanli Ouyang.
\newblock {FengWu}: {Pushing} the {Skillful} {Global} {Medium}-range {Weather}
  {Forecast} beyond 10 {Days} {Lead}, April 2023{\natexlab{b}}.
\newblock URL \url{http://arxiv.org/abs/2304.02948}.
\newblock arXiv:2304.02948 [physics].

\bibitem[Nguyen et~al.(2023)Nguyen, Shah, Bansal, Arcomano, Madireddy, Maulik,
  Kotamarthi, Foster, and Grover]{nguyen_scaling_2023}
Tung Nguyen, Rohan Shah, Hritik Bansal, Troy Arcomano, Sandeep Madireddy, Romit
  Maulik, Veerabhadra Kotamarthi, Ian Foster, and Aditya Grover.
\newblock Scaling transformer neural networks for skillful and reliable
  medium-range weather forecasting, December 2023.
\newblock URL \url{http://arxiv.org/abs/2312.03876}.
\newblock arXiv:2312.03876 [physics].

\bibitem[Hersbach et~al.(2020)Hersbach, Bell, Berrisford, Hirahara, Horányi,
  Muñoz-Sabater, Nicolas, Peubey, Radu, Schepers, Simmons, Soci, Abdalla,
  Abellan, Balsamo, Bechtold, Biavati, Bidlot, Bonavita, De~Chiara, Dahlgren,
  Dee, Diamantakis, Dragani, Flemming, Forbes, Fuentes, Geer, Haimberger,
  Healy, Hogan, Hólm, Janisková, Keeley, Laloyaux, Lopez, Lupu, Radnoti,
  de~Rosnay, Rozum, Vamborg, Villaume, and Thépaut]{hersbach_era5_2020}
Hans Hersbach, Bill Bell, Paul Berrisford, Shoji Hirahara, András Horányi,
  Joaquín Muñoz-Sabater, Julien Nicolas, Carole Peubey, Raluca Radu, Dinand
  Schepers, Adrian Simmons, Cornel Soci, Saleh Abdalla, Xavier Abellan,
  Gianpaolo Balsamo, Peter Bechtold, Gionata Biavati, Jean Bidlot, Massimo
  Bonavita, Giovanna De~Chiara, Per Dahlgren, Dick Dee, Michail Diamantakis,
  Rossana Dragani, Johannes Flemming, Richard Forbes, Manuel Fuentes, Alan
  Geer, Leo Haimberger, Sean Healy, Robin~J. Hogan, Elías Hólm, Marta
  Janisková, Sarah Keeley, Patrick Laloyaux, Philippe Lopez, Cristina Lupu,
  Gabor Radnoti, Patricia de~Rosnay, Iryna Rozum, Freja Vamborg, Sebastien
  Villaume, and Jean-Noël Thépaut.
\newblock The {ERA5} global reanalysis.
\newblock \emph{Quarterly Journal of the Royal Meteorological Society},
  146\penalty0 (730):\penalty0 1999--2049, 2020.
\newblock ISSN 1477-870X.
\newblock \doi{10.1002/qj.3803}.
\newblock URL \url{https://onlinelibrary.wiley.com/doi/abs/10.1002/qj.3803}.

\bibitem[Rasp et~al.(2020)Rasp, Dueben, Scher, Weyn, Mouatadid, and
  Thuerey]{rasp_weatherbench_2020}
Stephan Rasp, Peter~D. Dueben, Sebastian Scher, Jonathan~A. Weyn, Soukayna
  Mouatadid, and Nils Thuerey.
\newblock {WeatherBench}: {A} benchmark dataset for data-driven weather
  forecasting.
\newblock \emph{Journal of Advances in Modeling Earth Systems}, 12\penalty0
  (11), November 2020.
\newblock ISSN 1942-2466, 1942-2466.
\newblock \doi{10.1029/2020MS002203}.
\newblock URL \url{http://arxiv.org/abs/2002.00469}.
\newblock arXiv: 2002.00469.

\bibitem[Rasp et~al.(2023)Rasp, Hoyer, Merose, Langmore, Battaglia, Russel,
  Sanchez-Gonzalez, Yang, Carver, Agrawal, Chantry, Bouallegue, Dueben,
  Bromberg, Sisk, Barrington, Bell, and Sha]{rasp_weatherbench_2023}
Stephan Rasp, Stephan Hoyer, Alexander Merose, Ian Langmore, Peter Battaglia,
  Tyler Russel, Alvaro Sanchez-Gonzalez, Vivian Yang, Rob Carver, Shreya
  Agrawal, Matthew Chantry, Zied~Ben Bouallegue, Peter Dueben, Carla Bromberg,
  Jared Sisk, Luke Barrington, Aaron Bell, and Fei Sha.
\newblock {WeatherBench} 2: {A} benchmark for the next generation of
  data-driven global weather models, August 2023.
\newblock URL \url{http://arxiv.org/abs/2308.15560}.
\newblock arXiv:2308.15560 [physics].

\bibitem[Rasp and Thuerey(2021)]{rasp_data-driven_2021}
Stephan Rasp and Nils Thuerey.
\newblock Data-{Driven} {Medium}-{Range} {Weather} {Prediction} {With} a
  {Resnet} {Pretrained} on {Climate} {Simulations}: {A} {New} {Model} for
  {WeatherBench}.
\newblock \emph{Journal of Advances in Modeling Earth Systems}, 13\penalty0
  (2):\penalty0 e2020MS002405, 2021.
\newblock ISSN 1942-2466.
\newblock \doi{10.1029/2020MS002405}.
\newblock URL
  \url{https://onlinelibrary.wiley.com/doi/abs/10.1029/2020MS002405}.

\bibitem[Metz et~al.(2022)Metz, Freeman, Schoenholz, and
  Kachman]{metz_gradients_2022}
Luke Metz, C.~Daniel Freeman, Samuel~S. Schoenholz, and Tal Kachman.
\newblock Gradients are {Not} {All} {You} {Need}, January 2022.
\newblock URL \url{http://arxiv.org/abs/2111.05803}.
\newblock Number: arXiv:2111.05803 arXiv:2111.05803 [cs, stat].

\bibitem[Pascanu et~al.(2013)Pascanu, Mikolov, and
  Bengio]{mikhaeil_difficulty_2022}
Razvan Pascanu, Tomas Mikolov, and Yoshua Bengio.
\newblock On the difficulty of training recurrent neural networks.
\newblock In \emph{Proceedings of the 30th International Conference on
  International Conference on Machine Learning - Volume 28}, ICML'13, page
  III–1310–III–1318. JMLR.org, 2013.

\bibitem[Bengio et~al.(2015)Bengio, Vinyals, Jaitly, and
  Shazeer]{bengio_scheduled_2015}
Samy Bengio, Oriol Vinyals, Navdeep Jaitly, and Noam Shazeer.
\newblock Scheduled sampling for sequence prediction with recurrent neural
  networks.
\newblock In \emph{Proceedings of the 28th International Conference on Neural
  Information Processing Systems - Volume 1}, NIPS'15, pages 1171--1179,
  Cambridge, MA, USA, 2015. MIT Press.

\bibitem[Hu et~al.(2023)Hu, Chen, Wang, and Li]{hu_swinvrnn_2022}
Yuan Hu, Lei Chen, Zhibin Wang, and Hao Li.
\newblock Swinvrnn: A data-driven ensemble forecasting model via learned
  distribution perturbation.
\newblock \emph{Journal of Advances in Modeling Earth Systems}, 15\penalty0
  (2):\penalty0 e2022MS003211, 2023.
\newblock \doi{https://doi.org/10.1029/2022MS003211}.
\newblock e2022MS003211 2022MS003211.

\bibitem[Chen et~al.(2023{\natexlab{c}})Chen, Du, Hu, Wang, and
  Wang]{chen_swinrdm_2023}
Lei Chen, Fei Du, Yuan Hu, Fan Wang, and Zhibin Wang.
\newblock {SwinRDM}: {Integrate} {SwinRNN} with {Diffusion} {Model} towards
  {High}-{Resolution} and {High}-{Quality} {Weather} {Forecasting}.
\newblock January 2023{\natexlab{c}}.
\newblock \doi{10.48448/zn7f-fc64}.
\newblock URL \url{http://arxiv.org/abs/2306.03110}.
\newblock arXiv:2306.03110 [physics].

\bibitem[Ravuri et~al.(2021)Ravuri, Lenc, Willson, Kangin, Lam, Mirowski,
  Fitzsimons, Athanassiadou, Kashem, Madge, Prudden, Mandhane, Clark, Brock,
  Simonyan, Hadsell, Robinson, Clancy, Arribas, and
  Mohamed]{ravuri_skillful_2021}
Suman Ravuri, Karel Lenc, Matthew Willson, Dmitry Kangin, Remi Lam, Piotr
  Mirowski, Megan Fitzsimons, Maria Athanassiadou, Sheleem Kashem, Sam Madge,
  Rachel Prudden, Amol Mandhane, Aidan Clark, Andrew Brock, Karen Simonyan,
  Raia Hadsell, Niall Robinson, Ellen Clancy, Alberto Arribas, and Shakir
  Mohamed.
\newblock Skilful precipitation nowcasting using deep generative models of
  radar.
\newblock \emph{Nature}, 597\penalty0 (7878):\penalty0 672--677, September
  2021.
\newblock ISSN 1476-4687.
\newblock \doi{10.1038/s41586-021-03854-z}.
\newblock URL \url{https://doi.org/10.1038/s41586-021-03854-z}.

\bibitem[Leinonen et~al.(2021)Leinonen, Nerini, and
  Berne]{leinonen_stochastic_2021}
Jussi Leinonen, Daniele Nerini, and Alexis Berne.
\newblock Stochastic {Super}-{Resolution} for {Downscaling} {Time}-{Evolving}
  {Atmospheric} {Fields} with a {Generative} {Adversarial} {Network}.
\newblock \emph{IEEE Transactions on Geoscience and Remote Sensing},
  59\penalty0 (9):\penalty0 7211--7223, September 2021.
\newblock ISSN 0196-2892, 1558-0644.
\newblock \doi{10.1109/TGRS.2020.3032790}.
\newblock URL \url{http://arxiv.org/abs/2005.10374}.
\newblock arXiv:2005.10374 [physics, stat].

\bibitem[Adewoyin et~al.(2021)Adewoyin, Dueben, Watson, He, and
  Dutta]{adewoyin_tru-net_2021}
Rilwan~A. Adewoyin, Peter Dueben, Peter Watson, Yulan He, and Ritabrata Dutta.
\newblock {TRU}-{NET}: a deep learning approach to high resolution prediction
  of rainfall.
\newblock \emph{Machine Learning}, 110\penalty0 (8):\penalty0 2035--2062,
  August 2021.
\newblock ISSN 0885-6125, 1573-0565.
\newblock \doi{10.1007/s10994-021-06022-6}.

\bibitem[Sønderby et~al.(2020)Sønderby, Espeholt, Heek, Dehghani, Oliver,
  Salimans, Agrawal, Hickey, and Kalchbrenner]{sonderby_metnet_2020}
Casper~Kaae Sønderby, Lasse Espeholt, Jonathan Heek, Mostafa Dehghani, Avital
  Oliver, Tim Salimans, Shreya Agrawal, Jason Hickey, and Nal Kalchbrenner.
\newblock {MetNet}: {A} {Neural} {Weather} {Model} for {Precipitation}
  {Forecasting}.
\newblock \emph{arXiv:2003.12140 [physics, stat]}, March 2020.
\newblock URL \url{http://arxiv.org/abs/2003.12140}.
\newblock arXiv: 2003.12140.

\bibitem[Espeholt et~al.(2021)Espeholt, Agrawal, Sønderby, Kumar, Heek,
  Bromberg, Gazen, Hickey, Bell, and Kalchbrenner]{espeholt_skillful_2021}
Lasse Espeholt, Shreya Agrawal, Casper Sønderby, Manoj Kumar, Jonathan Heek,
  Carla Bromberg, Cenk Gazen, Jason Hickey, Aaron Bell, and Nal Kalchbrenner.
\newblock Skillful {Twelve} {Hour} {Precipitation} {Forecasts} using {Large}
  {Context} {Neural} {Networks}, November 2021.
\newblock URL \url{http://arxiv.org/abs/2111.07470}.
\newblock arXiv:2111.07470 [physics].

\bibitem[Palmer and Richardson(2014)]{palmer_decisions_2014}
Tim Palmer and David Richardson.
\newblock Decisions, decisions…!
\newblock \emph{ECMWF Newsletter}, pages 12--14, 2014 2014.
\newblock \doi{10.21957/bychj3cf}.

\bibitem[Palmer(2019)]{palmer_ecmwf_2019}
Tim Palmer.
\newblock The {ECMWF} ensemble prediction system: {Looking} back (more than) 25
  years and projecting forward 25 years.
\newblock \emph{Quarterly Journal of the Royal Meteorological Society},
  145\penalty0 (S1):\penalty0 12--24, 2019.
\newblock ISSN 1477-870X.
\newblock \doi{10.1002/qj.3383}.
\newblock URL \url{https://onlinelibrary.wiley.com/doi/abs/10.1002/qj.3383}.

\bibitem[Andrychowicz et~al.(2023)Andrychowicz, Espeholt, Li, Merchant, Merose,
  Zyda, Agrawal, and Kalchbrenner]{andrychowicz_deep_2023}
Marcin Andrychowicz, Lasse Espeholt, Di~Li, Samier Merchant, Alexander Merose,
  Fred Zyda, Shreya Agrawal, and Nal Kalchbrenner.
\newblock Deep {Learning} for {Day} {Forecasts} from {Sparse} {Observations},
  July 2023.
\newblock URL \url{http://arxiv.org/abs/2306.06079}.
\newblock arXiv:2306.06079 [physics].

\bibitem[Lessig et~al.(2023)Lessig, Luise, Gong, Langguth, Stadler, and
  Schultz]{lessig_atmorep_2023}
Christian Lessig, Ilaria Luise, Bing Gong, Michael Langguth, Scarlet Stadler,
  and Martin Schultz.
\newblock {AtmoRep}: {A} stochastic model of atmosphere dynamics using large
  scale representation learning, August 2023.
\newblock URL \url{http://arxiv.org/abs/2308.13280}.
\newblock arXiv:2308.13280 [physics].

\bibitem[Bond-Taylor et~al.(2021)Bond-Taylor, Leach, Long, and
  Willcocks]{bond-taylor_deep_2021}
Sam Bond-Taylor, Adam Leach, Yang Long, and Chris~G. Willcocks.
\newblock Deep {Generative} {Modelling}: {A} {Comparative} {Review} of {VAEs},
  {GANs}, {Normalizing} {Flows}, {Energy}-{Based} and {Autoregressive}
  {Models}.
\newblock \emph{IEEE Transactions on Pattern Analysis and Machine
  Intelligence}, pages 1--1, 2021.
\newblock ISSN 0162-8828, 2160-9292, 1939-3539.
\newblock \doi{10.1109/TPAMI.2021.3116668}.
\newblock URL \url{http://arxiv.org/abs/2103.04922}.
\newblock arXiv: 2103.04922.

\bibitem[Rezende et~al.(2014)Rezende, Mohamed, and
  Wierstra]{rezende_stochastic_2014}
Danilo~Jimenez Rezende, Shakir Mohamed, and Daan Wierstra.
\newblock Stochastic backpropagation and approximate inference in deep
  generative models.
\newblock In \emph{Proceedings of the 31st International Conference on
  International Conference on Machine Learning - Volume 32}, ICML'14, page
  II–1278–II–1286. JMLR.org, 2014.

\bibitem[Kingma and Welling(2014)]{kingma_auto-encoding_2014}
Diederik~P. Kingma and Max Welling.
\newblock Auto-encoding variational bayes.
\newblock In \emph{International Conference on Learning Representations}, 2014.

\bibitem[Chen et~al.(2023{\natexlab{d}})Chen, Zhong, Wu, Chen, Xie, Chao, Lin,
  Hu, Lu, Li, and Qi]{chen_fuxi-s2s_2023}
Lei Chen, Xiaohui Zhong, Jie Wu, Deliang Chen, Shangping Xie, Qingchen Chao,
  Chensen Lin, Zixin Hu, Bo~Lu, Hao Li, and Yuan Qi.
\newblock {FuXi}-{S2S}: {An} accurate machine learning model for global
  subseasonal forecasts, December 2023{\natexlab{d}}.
\newblock URL \url{http://arxiv.org/abs/2312.09926}.
\newblock arXiv:2312.09926 [physics].

\bibitem[Babaeizadeh et~al.(2018)Babaeizadeh, Finn, Erhan, Campbell, and
  Levine]{babaeizadeh_stochastic_2018}
Mohammad Babaeizadeh, Chelsea Finn, Dumitru Erhan, Roy~H. Campbell, and Sergey
  Levine.
\newblock Stochastic variational video prediction.
\newblock In \emph{International Conference on Learning Representations}, 2018.

\bibitem[Denton and Fergus(2018)]{denton_stochastic_2018}
Emily Denton and Rob Fergus.
\newblock Stochastic video generation with a learned prior.
\newblock In Jennifer Dy and Andreas Krause, editors, \emph{Proceedings of the
  35th International Conference on Machine Learning}, volume~80 of
  \emph{Proceedings of Machine Learning Research}, pages 1174--1183. PMLR,
  10--15 Jul 2018.

\bibitem[Goodfellow et~al.(2014)Goodfellow, Pouget-Abadie, Mirza, Xu,
  Warde-Farley, Ozair, Courville, and Bengio]{goodfellow_generative_2014}
Ian Goodfellow, Jean Pouget-Abadie, Mehdi Mirza, Bing Xu, David Warde-Farley,
  Sherjil Ozair, Aaron Courville, and Yoshua Bengio.
\newblock Generative adversarial nets.
\newblock In Z.~Ghahramani, M.~Welling, C.~Cortes, N.~Lawrence, and K.Q.
  Weinberger, editors, \emph{Advances in Neural Information Processing
  Systems}, volume~27. Curran Associates, Inc., 2014.

\bibitem[Gong et~al.(2022)Gong, Langguth, Ji, Mozaffari, Stadtler, Mache, and
  Schultz]{gong_temperature_2022}
Bing Gong, Michael Langguth, Yan Ji, Amirpasha Mozaffari, Scarlet Stadtler,
  Karim Mache, and Martin~G. Schultz.
\newblock Temperature forecasting by deep learning methods.
\newblock preprint, Climate and Earth system modeling, March 2022.
\newblock URL \url{https://gmd.copernicus.org/preprints/gmd-2021-430/}.

\bibitem[Hoffmann and Lessig(2023)]{hoffmann_atmodist_2022}
Sebastian Hoffmann and Christian Lessig.
\newblock Atmodist: Self-supervised representation learning for atmospheric
  dynamics.
\newblock \emph{Environmental Data Science}, 2:\penalty0 e6, 2023.
\newblock \doi{10.1017/eds.2023.1}.

\bibitem[Ho et~al.(2020)Ho, Jain, and Abbeel]{ho_denoising_2020}
Jonathan Ho, Ajay Jain, and Pieter Abbeel.
\newblock Denoising {Diffusion} {Probabilistic} {Models}.
\newblock \emph{arXiv:2006.11239 [cs, stat]}, December 2020.
\newblock URL \url{http://arxiv.org/abs/2006.11239}.
\newblock arXiv: 2006.11239.

\bibitem[Sohl-Dickstein et~al.(2015)Sohl-Dickstein, Weiss, Maheswaranathan, and
  Ganguli]{sohl-dickstein_deep_2015}
Jascha Sohl-Dickstein, Eric~A. Weiss, Niru Maheswaranathan, and Surya Ganguli.
\newblock Deep {Unsupervised} {Learning} using {Nonequilibrium}
  {Thermodynamics}.
\newblock \emph{arXiv:1503.03585 [cond-mat, q-bio, stat]}, November 2015.
\newblock URL \url{http://arxiv.org/abs/1503.03585}.
\newblock arXiv: 1503.03585.

\bibitem[Song et~al.(2021)Song, Meng, and Ermon]{song_denoising_2021}
Jiaming Song, Chenlin Meng, and Stefano Ermon.
\newblock Denoising {Diffusion} {Implicit} {Models}.
\newblock \emph{arXiv:2010.02502 [cs]}, November 2021.
\newblock URL \url{http://arxiv.org/abs/2010.02502}.
\newblock arXiv: 2010.02502.

\bibitem[Price et~al.(2023)Price, Sanchez-Gonzalez, Alet, Ewalds, El-Kadi,
  Stott, Mohamed, Battaglia, Lam, and Willson]{price_gencast_2023}
Ilan Price, Alvaro Sanchez-Gonzalez, Ferran Alet, Timo Ewalds, Andrew El-Kadi,
  Jacklynn Stott, Shakir Mohamed, Peter Battaglia, Remi Lam, and Matthew
  Willson.
\newblock {GenCast}: {Diffusion}-based ensemble forecasting for medium-range
  weather, December 2023.
\newblock URL \url{http://arxiv.org/abs/2312.15796}.
\newblock arXiv:2312.15796 [physics].

\bibitem[Cachay et~al.(2023)Cachay, Zhao, Joren, and Yu]{cachay_dyffusion_2023}
Salva~Rühling Cachay, Bo~Zhao, Hailey Joren, and Rose Yu.
\newblock {DYffusion}: {A} {Dynamics}-informed {Diffusion} {Model} for
  {Spatiotemporal} {Forecasting}, October 2023.
\newblock URL \url{http://arxiv.org/abs/2306.01984}.
\newblock arXiv:2306.01984 [cs, stat].

\bibitem[Li et~al.(2023)Li, Carver, Lopez-Gomez, Sha, and
  Anderson]{li_seeds_2023}
Lizao Li, Rob Carver, Ignacio Lopez-Gomez, Fei Sha, and John Anderson.
\newblock {SEEDS}: {Emulation} of {Weather} {Forecast} {Ensembles} with
  {Diffusion} {Models}, October 2023.
\newblock URL \url{http://arxiv.org/abs/2306.14066}.
\newblock arXiv:2306.14066 [physics].

\bibitem[Leinonen et~al.(2023)Leinonen, Hamann, Nerini, Germann, and
  Franch]{leinonen_latent_2023}
Jussi Leinonen, Ulrich Hamann, Daniele Nerini, Urs Germann, and Gabriele
  Franch.
\newblock Latent diffusion models for generative precipitation nowcasting with
  accurate uncertainty quantification, April 2023.
\newblock URL \url{http://arxiv.org/abs/2304.12891}.
\newblock arXiv:2304.12891 [physics].

\bibitem[Graubner et~al.(2022)Graubner, Kamyar~Azizzadenesheli, Pathak,
  Mardani, Pritchard, Kashinath, and Anandkumar]{graubner_calibration_2022}
Andre Graubner, Kamyar Kamyar~Azizzadenesheli, Jaideep Pathak, Morteza Mardani,
  Mike Pritchard, Karthik Kashinath, and Anima Anandkumar.
\newblock Calibration of large neural weather models.
\newblock In \emph{NeurIPS 2022 Workshop on Tackling Climate Change with
  Machine Learning}, 2022.

\bibitem[Kendall and Gal(2017)]{kendall_what_2017}
Alex Kendall and Yarin Gal.
\newblock What uncertainties do we need in bayesian deep learning for computer
  vision?
\newblock In \emph{Advances in Neural Information Processing Systems},
  volume~30. Curran Associates, Inc., 2017.

\bibitem[Stirn and Knowles(2020)]{stirn_variational_2020}
Andrew Stirn and David~A. Knowles.
\newblock Variational {Variance}: {Simple}, {Reliable}, {Calibrated}
  {Heteroscedastic} {Noise} {Variance} {Parameterization}, October 2020.
\newblock URL \url{http://arxiv.org/abs/2006.04910}.
\newblock arXiv:2006.04910 [cs, stat].

\bibitem[Gneiting and Raftery(2007)]{gneiting_strictly_2007}
Tilmann Gneiting and Adrian~E Raftery.
\newblock Strictly proper scoring rules, prediction, and estimation.
\newblock \emph{Journal of the American Statistical Association}, 102\penalty0
  (477):\penalty0 359--378, 2007.
\newblock \doi{10.1198/016214506000001437}.

\bibitem[Seitzer et~al.(2022)Seitzer, Tavakoli, Antic, and
  Martius]{seitzer_pitfalls_2022}
Maximilian Seitzer, Arash Tavakoli, Dimitrije Antic, and Georg Martius.
\newblock On the pitfalls of heteroscedastic uncertainty estimation with
  probabilistic neural networks.
\newblock In \emph{International Conference on Learning Representations}, 2022.

\bibitem[Berrisch and Ziel(2021)]{berrisch_crps_2021}
Jonathan Berrisch and Florian Ziel.
\newblock Crps learning.
\newblock \emph{Journal of Econometrics}, 2021.
\newblock ISSN 0304-4076.
\newblock \doi{https://doi.org/10.1016/j.jeconom.2021.11.008}.

\bibitem[Kochkov et~al.(2023)Kochkov, Yuval, Langmore, Norgaard, Smith, Mooers,
  Lottes, Rasp, Düben, Klöwer, Hatfield, Battaglia, Sanchez-Gonzalez,
  Willson, Brenner, and Hoyer]{kochkov_neural_2023}
Dmitrii Kochkov, Janni Yuval, Ian Langmore, Peter Norgaard, Jamie Smith,
  Griffin Mooers, James Lottes, Stephan Rasp, Peter Düben, Milan Klöwer, Sam
  Hatfield, Peter Battaglia, Alvaro Sanchez-Gonzalez, Matthew Willson,
  Michael~P. Brenner, and Stephan Hoyer.
\newblock Neural {General} {Circulation} {Models}, November 2023.
\newblock URL \url{http://arxiv.org/abs/2311.07222}.
\newblock arXiv:2311.07222 [physics].

\bibitem[Chapman et~al.(2022)Chapman, Monache, Alessandrini, Subramanian,
  Ralph, Xie, Lerch, and Hayatbini]{chapman_probabilistic_2022}
William~E. Chapman, Luca~Delle Monache, Stefano Alessandrini, Aneesh~C.
  Subramanian, F.~Martin Ralph, Shang-Ping Xie, Sebastian Lerch, and Negin
  Hayatbini.
\newblock Probabilistic predictions from deterministic atmospheric river
  forecasts with deep learning.
\newblock \emph{Monthly Weather Review}, 150\penalty0 (1):\penalty0 215--234,
  2022.
\newblock \doi{https://doi.org/10.1175/MWR-D-21-0106.1}.

\bibitem[He et~al.(2016)He, Zhang, Ren, and Sun]{he_deep_2015}
Kaiming He, Xiangyu Zhang, Shaoqing Ren, and Jian Sun.
\newblock Deep residual learning for image recognition.
\newblock In \emph{2016 IEEE Conference on Computer Vision and Pattern
  Recognition (CVPR)}, pages 770--778, 2016.
\newblock \doi{10.1109/CVPR.2016.90}.

\bibitem[He et~al.(2015)He, Zhang, Ren, and Sun]{he_delving_2015}
Kaiming He, Xiangyu Zhang, Shaoqing Ren, and Jian Sun.
\newblock Delving {Deep} into {Rectifiers}: {Surpassing} {Human}-{Level}
  {Performance} on {ImageNet} {Classification}.
\newblock In \emph{2015 {IEEE} {International} {Conference} on {Computer}
  {Vision} ({ICCV})}, pages 1026--1034, Santiago, Chile, December 2015. IEEE.
\newblock ISBN 978-1-4673-8391-2.
\newblock \doi{10.1109/ICCV.2015.123}.
\newblock URL \url{http://ieeexplore.ieee.org/document/7410480/}.

\bibitem[Ioffe and Szegedy(2015)]{ioffe_batch_2015}
Sergey Ioffe and Christian Szegedy.
\newblock Batch normalization: Accelerating deep network training by reducing
  internal covariate shift.
\newblock In \emph{Proceedings of the 32nd International Conference on
  International Conference on Machine Learning - Volume 37}, ICML'15, page
  448–456. JMLR.org, 2015.

\bibitem[Ba et~al.(2016)Ba, Kiros, and Hinton]{ba_layer_2016}
Jimmy~Lei Ba, Jamie~Ryan Kiros, and Geoffrey~E. Hinton.
\newblock Layer {Normalization}.
\newblock \emph{arXiv:1607.06450 [cs, stat]}, July 2016.
\newblock URL \url{http://arxiv.org/abs/1607.06450}.
\newblock arXiv: 1607.06450.

\bibitem[Ulyanov et~al.(2017)Ulyanov, Vedaldi, and
  Lempitsky]{ulyanov_instance_2017}
Dmitry Ulyanov, Andrea Vedaldi, and Victor Lempitsky.
\newblock Instance {Normalization}: {The} {Missing} {Ingredient} for {Fast}
  {Stylization}, November 2017.
\newblock URL \url{http://arxiv.org/abs/1607.08022}.
\newblock arXiv:1607.08022 [cs].

\bibitem[Wu and He(2020)]{wu_group_2018}
Yuxin Wu and Kaiming He.
\newblock Group {Normalization}.
\newblock \emph{International Journal of Computer Vision}, 128\penalty0
  (3):\penalty0 742--755, March 2020.
\newblock ISSN 0920-5691, 1573-1405.
\newblock \doi{10.1007/s11263-019-01198-w}.

\bibitem[Lecun and Bengio(1995)]{lecun_convolutional_1995}
Yann Lecun and Yoshua Bengio.
\newblock Convolutional {Networks} for {Images}, {Speech} and {Time} {Series}.
\newblock In Michael~A. Arbib, editor, \emph{The {Handbook} of {Brain} {Theory}
  and {Neural} {Networks}}, pages 255--258. The MIT Press, 1995.

\bibitem[LeCun et~al.(1998)LeCun, Bottou, Bengio, and
  Haffner]{lecun_gradient-based_1998}
Yann LeCun, Léon Bottou, Yoshua Bengio, and Patrick Haffner.
\newblock Gradient-{Based} {Learning} {Applied} to {Document} {Recognition}.
\newblock In \emph{Proceedings of the {IEEE}}, volume~86, pages 2278--2324,
  1998.
\newblock URL
  \url{http://citeseerx.ist.psu.edu/viewdoc/summary?doi=10.1.1.42.7665}.
\newblock Issue: 11.

\bibitem[Bronstein et~al.(2021)Bronstein, Bruna, Cohen, and
  Veličković]{bronstein_geometric_2021}
Michael~M. Bronstein, Joan Bruna, Taco Cohen, and Petar Veličković.
\newblock Geometric {Deep} {Learning}: {Grids}, {Groups}, {Graphs},
  {Geodesics}, and {Gauges}, May 2021.
\newblock URL \url{http://arxiv.org/abs/2104.13478}.
\newblock arXiv:2104.13478 [cs, stat].

\bibitem[Yu and Koltun(2016)]{yu_multi-scale_2016}
Fisher Yu and Vladlen Koltun.
\newblock Multi-scale context aggregation by dilated convolutions.
\newblock In \emph{International Conference on Learning Representations}, 2016.

\bibitem[Chollet(2017)]{chollet_xception_2017}
François Chollet.
\newblock Xception: {Deep} {Learning} with {Depthwise} {Separable}
  {Convolutions}.
\newblock \emph{arXiv:1610.02357 [cs]}, April 2017.
\newblock URL \url{http://arxiv.org/abs/1610.02357}.
\newblock arXiv: 1610.02357.

\bibitem[Liu et~al.(2022)Liu, Mao, Wu, Feichtenhofer, Darrell, and
  Xie]{liu_convnet_2022}
Zhuang Liu, Hanzi Mao, Chao-Yuan Wu, Christoph Feichtenhofer, Trevor Darrell,
  and Saining Xie.
\newblock A convnet for the 2020s.
\newblock In \emph{2022 IEEE/CVF Conference on Computer Vision and Pattern
  Recognition (CVPR)}, pages 11966--11976, 2022.
\newblock \doi{10.1109/CVPR52688.2022.01167}.

\bibitem[Gilmer et~al.(2017)Gilmer, Schoenholz, Riley, Vinyals, and
  Dahl]{gilmer_neural_2017}
Justin Gilmer, Samuel~S. Schoenholz, Patrick~F. Riley, Oriol Vinyals, and
  George~E. Dahl.
\newblock Neural message passing for quantum chemistry.
\newblock In Doina Precup and Yee~Whye Teh, editors, \emph{Proceedings of the
  34th International Conference on Machine Learning}, volume~70 of
  \emph{Proceedings of Machine Learning Research}, pages 1263--1272. PMLR,
  06--11 Aug 2017.

\bibitem[Vaswani et~al.(2017)Vaswani, Shazeer, Parmar, Uszkoreit, Jones, Gomez,
  Kaiser, and Polosukhin]{vaswani_attention_2017}
Ashish Vaswani, Noam Shazeer, Niki Parmar, Jakob Uszkoreit, Llion Jones,
  Aidan~N Gomez, \L~ukasz Kaiser, and Illia Polosukhin.
\newblock Attention is all you need.
\newblock In \emph{Advances in Neural Information Processing Systems},
  volume~30. Curran Associates, Inc., 2017.

\bibitem[Tsai et~al.(2019)Tsai, Bai, Yamada, Morency, and
  Salakhutdinov]{tsai_transformer_2019}
Yao-Hung~Hubert Tsai, Shaojie Bai, Makoto Yamada, Louis-Philippe Morency, and
  Ruslan Salakhutdinov.
\newblock Transformer dissection: An unified understanding for transformer{'}s
  attention via the lens of kernel.
\newblock In \emph{Proceedings of the 2019 Conference on Empirical Methods in
  Natural Language Processing and the 9th International Joint Conference on
  Natural Language Processing (EMNLP-IJCNLP)}, pages 4344--4353, Hong Kong,
  China, November 2019. Association for Computational Linguistics.
\newblock \doi{10.18653/v1/D19-1443}.

\bibitem[Li et~al.(2022)Li, Huang, Liu, and Anandkumar]{li_fourier_2022}
Zongyi Li, Daniel~Zhengyu Huang, Burigede Liu, and Anima Anandkumar.
\newblock Fourier {Neural} {Operator} with {Learned} {Deformations} for {PDEs}
  on {General} {Geometries}, July 2022.
\newblock URL \url{http://arxiv.org/abs/2207.05209}.
\newblock arXiv:2207.05209 [cs, math].

\bibitem[Guibas et~al.(2022)Guibas, Mardani, Li, Tao, Anandkumar, and
  Catanzaro]{guibas_adaptive_2022}
John Guibas, Morteza Mardani, Zongyi Li, Andrew Tao, Anima Anandkumar, and
  Bryan Catanzaro.
\newblock Efficient token mixing for transformers via adaptive fourier neural
  operators.
\newblock In \emph{International Conference on Learning Representations}, 2022.

\bibitem[Dosovitskiy et~al.(2021)Dosovitskiy, Beyer, Kolesnikov, Weissenborn,
  Zhai, Unterthiner, Dehghani, Minderer, Heigold, Gelly, Uszkoreit, and
  Houlsby]{dosovitskiy_image_2021}
Alexey Dosovitskiy, Lucas Beyer, Alexander Kolesnikov, Dirk Weissenborn,
  Xiaohua Zhai, Thomas Unterthiner, Mostafa Dehghani, Matthias Minderer, Georg
  Heigold, Sylvain Gelly, Jakob Uszkoreit, and Neil Houlsby.
\newblock An image is worth 16x16 words: Transformers for image recognition at
  scale.
\newblock In \emph{International Conference on Learning Representations}, 2021.

\bibitem[Sanchez-Gonzalez et~al.(2018)Sanchez-Gonzalez, Heess, Springenberg,
  Merel, Riedmiller, Hadsell, and Battaglia]{sanchez-gonzalez_graph_2018}
Alvaro Sanchez-Gonzalez, Nicolas Heess, Jost~Tobias Springenberg, Josh Merel,
  Martin Riedmiller, Raia Hadsell, and Peter Battaglia.
\newblock Graph networks as learnable physics engines for inference and
  control.
\newblock \emph{arXiv:1806.01242 [cs, stat]}, June 2018.
\newblock URL \url{http://arxiv.org/abs/1806.01242}.
\newblock arXiv: 1806.01242.

\bibitem[Liu et~al.(2021)Liu, Lin, Cao, Hu, Wei, Zhang, Lin, and
  Guo]{liu_swin_2021}
Z.~Liu, Y.~Lin, Y.~Cao, H.~Hu, Y.~Wei, Z.~Zhang, S.~Lin, and B.~Guo.
\newblock Swin transformer: Hierarchical vision transformer using shifted
  windows.
\newblock In \emph{2021 IEEE/CVF International Conference on Computer Vision
  (ICCV)}, pages 9992--10002, Los Alamitos, CA, USA, oct 2021. IEEE Computer
  Society.
\newblock \doi{10.1109/ICCV48922.2021.00986}.

\bibitem[Ronneberger et~al.(2015)Ronneberger, Fischer, and
  Brox]{ronneberger_u-net_2015}
Olaf Ronneberger, Philipp Fischer, and Thomas Brox.
\newblock U-net: Convolutional networks for biomedical image segmentation.
\newblock In Nassir Navab, Joachim Hornegger, William~M. Wells, and
  Alejandro~F. Frangi, editors, \emph{Medical Image Computing and
  Computer-Assisted Intervention -- MICCAI 2015}, pages 234--241, Cham, 2015.
  Springer International Publishing.
\newblock ISBN 978-3-319-24574-4.

\bibitem[Ebert-Uphoff and Hilburn(2020)]{ebert-uphoff_evaluation_2020}
Imme Ebert-Uphoff and Kyle~A. Hilburn.
\newblock Evaluation, {Tuning} and {Interpretation} of {Neural} {Networks} for
  {Meteorological} {Applications}, May 2020.
\newblock URL \url{http://arxiv.org/abs/2005.03126}.
\newblock arXiv:2005.03126 [physics].

\bibitem[Lagerquist et~al.(2021)Lagerquist, Turner, Ebert-Uphoff, Stewart, and
  Hagerty]{lagerquist_using_2021}
Ryan Lagerquist, David Turner, Imme Ebert-Uphoff, Jebb Stewart, and Venita
  Hagerty.
\newblock Using {Deep} {Learning} to {Emulate} and {Accelerate} a {Radiative}
  {Transfer} {Model}.
\newblock \emph{Journal of Atmospheric and Oceanic Technology}, 38\penalty0
  (10):\penalty0 1673--1696, October 2021.
\newblock ISSN 0739-0572, 1520-0426.
\newblock \doi{10.1175/JTECH-D-21-0007.1}.
\newblock URL
  \url{https://journals.ametsoc.org/view/journals/atot/38/10/JTECH-D-21-0007.1.xml}.
\newblock Publisher: American Meteorological Society Section: Journal of
  Atmospheric and Oceanic Technology.

\bibitem[Jaegle et~al.(2022)Jaegle, Borgeaud, Alayrac, Doersch, Ionescu, Ding,
  Koppula, Zoran, Brock, Shelhamer, Hénaff, Botvinick, Zisserman, Vinyals, and
  Carreira]{jaegle_perceiver_2022}
Andrew Jaegle, Sebastian Borgeaud, Jean-Baptiste Alayrac, Carl Doersch, Catalin
  Ionescu, David Ding, Skanda Koppula, Daniel Zoran, Andrew Brock, Evan
  Shelhamer, Olivier Hénaff, Matthew~M. Botvinick, Andrew Zisserman, Oriol
  Vinyals, and Joāo Carreira.
\newblock Perceiver {IO}: {A} {General} {Architecture} for {Structured}
  {Inputs} \& {Outputs}, March 2022.
\newblock URL \url{http://arxiv.org/abs/2107.14795}.
\newblock arXiv:2107.14795 [cs, eess].

\bibitem[de~Bezenac et~al.(2018)de~Bezenac, Pajot, and
  Gallinari]{de_bezenac_deep_2018}
Emmanuel de~Bezenac, Arthur Pajot, and Patrick Gallinari.
\newblock Deep {Learning} for {Physical} {Processes}: {Incorporating} {Prior}
  {Scientific} {Knowledge}.
\newblock \emph{arXiv:1711.07970 [cs, stat]}, January 2018.
\newblock URL \url{http://arxiv.org/abs/1711.07970}.
\newblock arXiv: 1711.07970.

\bibitem[van Amersfoort et~al.(2017)van Amersfoort, Kannan, Ranzato, Szlam,
  Tran, and Chintala]{van_amersfoort_transformation-based_2017}
Joost van Amersfoort, Anitha Kannan, Marc'Aurelio Ranzato, Arthur Szlam,
  Du~Tran, and Soumith Chintala.
\newblock Transformation-{Based} {Models} of {Video} {Sequences}.
\newblock \emph{arXiv:1701.08435 [cs]}, April 2017.
\newblock URL \url{http://arxiv.org/abs/1701.08435}.
\newblock arXiv: 1701.08435.

\bibitem[Luc et~al.(2020)Luc, Clark, Dieleman, Casas, Doron, Cassirer, and
  Simonyan]{luc_transformation-based_2020}
Pauline Luc, Aidan Clark, Sander Dieleman, Diego de~Las Casas, Yotam Doron,
  Albin Cassirer, and Karen Simonyan.
\newblock Transformation-based {Adversarial} {Video} {Prediction} on
  {Large}-{Scale} {Data}.
\newblock \emph{arXiv:2003.04035 [cs]}, December 2020.
\newblock URL \url{http://arxiv.org/abs/2003.04035}.
\newblock arXiv: 2003.04035.

\bibitem[Robbins and Monro(1951)]{robbins_stochastic_1951}
Herbert Robbins and Sutton Monro.
\newblock A {Stochastic} {Approximation} {Method}.
\newblock \emph{The Annals of Mathematical Statistics}, 22\penalty0
  (3):\penalty0 400--407, September 1951.
\newblock ISSN 0003-4851, 2168-8990.
\newblock \doi{10.1214/aoms/1177729586}.
\newblock URL
  \url{https://projecteuclid.org/journals/annals-of-mathematical-statistics/volume-22/issue-3/A-Stochastic-Approximation-Method/10.1214/aoms/1177729586.full}.
\newblock Publisher: Institute of Mathematical Statistics.

\bibitem[Bartlett et~al.(2021)Bartlett, Montanari, and
  Rakhlin]{bartlett_deep_2021}
Peter~L. Bartlett, Andrea Montanari, and Alexander Rakhlin.
\newblock Deep learning: a statistical viewpoint.
\newblock \emph{Acta Numerica}, 30:\penalty0 87--201, May 2021.
\newblock ISSN 0962-4929, 1474-0508.
\newblock \doi{10.1017/S0962492921000027}.
\newblock URL
  \url{https://www.cambridge.org/core/product/identifier/S0962492921000027/type/journal_article}.

\bibitem[Hochreiter and Schmidhuber(1997)]{hochreiter_flat_1997}
Sepp Hochreiter and Jürgen Schmidhuber.
\newblock Flat {Minima}.
\newblock \emph{Neural Computation}, 9\penalty0 (1):\penalty0 1--42, January
  1997.
\newblock ISSN 0899-7667.
\newblock \doi{10.1162/neco.1997.9.1.1}.
\newblock URL \url{https://doi.org/10.1162/neco.1997.9.1.1}.

\bibitem[Keskar et~al.(2017)Keskar, Mudigere, Nocedal, Smelyanskiy, and
  Tang]{keskar_large-batch_2017}
Nitish~Shirish Keskar, Dheevatsa Mudigere, Jorge Nocedal, Mikhail Smelyanskiy,
  and Ping Tak~Peter Tang.
\newblock On large-batch training for deep learning: Generalization gap and
  sharp minima.
\newblock In \emph{International Conference on Learning Representations}, 2017.

\bibitem[Xing et~al.(2018)Xing, Arpit, Tsirigotis, and Bengio]{xing_walk_2018}
Chen Xing, Devansh Arpit, Christos Tsirigotis, and Yoshua Bengio.
\newblock A {Walk} with {SGD}.
\newblock \emph{arXiv:1802.08770 [cs, stat]}, May 2018.
\newblock URL \url{http://arxiv.org/abs/1802.08770}.
\newblock arXiv: 1802.08770.

\bibitem[Goyal et~al.(2018)Goyal, Dollár, Girshick, Noordhuis, Wesolowski,
  Kyrola, Tulloch, Jia, and He]{goyal_accurate_2018}
Priya Goyal, Piotr Dollár, Ross Girshick, Pieter Noordhuis, Lukasz Wesolowski,
  Aapo Kyrola, Andrew Tulloch, Yangqing Jia, and Kaiming He.
\newblock Accurate, {Large} {Minibatch} {SGD}: {Training} {ImageNet} in 1
  {Hour}, April 2018.
\newblock URL \url{http://arxiv.org/abs/1706.02677}.
\newblock Number: arXiv:1706.02677 arXiv:1706.02677 [cs].

\bibitem[Sutskever et~al.(2013)Sutskever, Martens, Dahl, and
  Hinton]{sutskever_importance_2013}
Ilya Sutskever, James Martens, George Dahl, and Geoffrey Hinton.
\newblock On the importance of initialization and momentum in deep learning.
\newblock In \emph{Proceedings of the 30th International Conference on
  International Conference on Machine Learning - Volume 28}, ICML'13, page
  III–1139–III–1147. JMLR.org, 2013.

\bibitem[Duchi et~al.(2011)Duchi, Hazan, and Singer]{duchi_adaptive_2011}
John Duchi, Elad Hazan, and Yoram Singer.
\newblock Adaptive {Subgradient} {Methods} for {Online} {Learning} and
  {Stochastic} {Optimization}.
\newblock 2011.

\bibitem[Kingma and Ba(2015)]{kingma_adam_2017}
Diederik~P. Kingma and Jimmy Ba.
\newblock Adam: {A} {Method} for {Stochastic} {Optimization}.
\newblock In \emph{International Conference on Learning Representations}, 2015.

\bibitem[Loshchilov and Hutter(2019)]{loshchilov_decoupled_2019}
Ilya Loshchilov and Frank Hutter.
\newblock Decoupled weight decay regularization.
\newblock In \emph{International Conference on Learning Representations}, 2019.

\bibitem[Hanson and Pratt(1988)]{hanson_comparing_nodate}
Stephen~Jose Hanson and Lorien~Y Pratt.
\newblock Comparing {Biases} for {Minimal} {Network} {Construction} with
  {Back}-{Propagation}.
\newblock 1988.

\bibitem[Srivastava et~al.(2014)Srivastava, Hinton, Krizhevsky, Sutskever, and
  Salakhutdinov]{srivastava_dropout_2014}
Nitish Srivastava, Geoffrey Hinton, Alex Krizhevsky, Ilya Sutskever, and Ruslan
  Salakhutdinov.
\newblock Dropout: A simple way to prevent neural networks from overfitting.
\newblock \emph{Journal of Machine Learning Research}, 15\penalty0
  (56):\penalty0 1929--1958, 2014.

\bibitem[Ghiasi et~al.(2018)Ghiasi, Lin, and Le]{ghiasi_dropblock_2018}
Golnaz Ghiasi, Tsung-Yi Lin, and Quoc~V Le.
\newblock Dropblock: A regularization method for convolutional networks.
\newblock In S.~Bengio, H.~Wallach, H.~Larochelle, K.~Grauman, N.~Cesa-Bianchi,
  and R.~Garnett, editors, \emph{Advances in Neural Information Processing
  Systems}, volume~31. Curran Associates, Inc., 2018.

\bibitem[Wu et~al.(2018)Wu, Nagarajan, Kumar, Rennie, Davis, Grauman, and
  Feris]{wu_blockdrop_2019}
Zuxuan Wu, Tushar Nagarajan, Abhishek Kumar, Steven Rennie, Larry~S. Davis,
  Kristen Grauman, and Rog{\'{e}}rio~Schmidt Feris.
\newblock Blockdrop: Dynamic inference paths in residual networks.
\newblock In \emph{2018 {IEEE} Conference on Computer Vision and Pattern
  Recognition, {CVPR} 2018, Salt Lake City, UT, USA, June 18-22, 2018}, pages
  8817--8826. Computer Vision Foundation / {IEEE} Computer Society, 2018.
\newblock \doi{10.1109/CVPR.2018.00919}.

\bibitem[He et~al.(2021)He, Chen, Xie, Li, Dollár, and
  Girshick]{he_masked_2021}
Kaiming He, Xinlei Chen, Saining Xie, Yanghao Li, Piotr Dollár, and Ross
  Girshick.
\newblock Masked {Autoencoders} {Are} {Scalable} {Vision} {Learners}, December
  2021.
\newblock URL \url{http://arxiv.org/abs/2111.06377}.
\newblock arXiv:2111.06377 [cs].

\bibitem[Bengio et~al.(2009)Bengio, Louradour, Collobert, and
  Weston]{bengio_curriculum_2009}
Yoshua Bengio, Jérôme Louradour, Ronan Collobert, and Jason Weston.
\newblock Curriculum learning.
\newblock In \emph{Proceedings of the 26th {Annual} {International}
  {Conference} on {Machine} {Learning}}, {ICML} '09, pages 41--48, New York,
  NY, USA, June 2009. Association for Computing Machinery.
\newblock ISBN 978-1-60558-516-1.
\newblock \doi{10.1145/1553374.1553380}.
\newblock URL \url{https://doi.org/10.1145/1553374.1553380}.

\bibitem[Erhan et~al.(2009)Erhan, Manzagol, Bengio, Bengio, and
  Vincent]{erhan_difficulty_2009}
Dumitru Erhan, Pierre-Antoine Manzagol, Yoshua Bengio, Samy Bengio, and Pascal
  Vincent.
\newblock The {Difficulty} of {Training} {Deep} {Architectures} and the
  {Effect} of {Unsupervised} {Pre}-{Training}.
\newblock In \emph{Proceedings of the {Twelth} {International} {Conference} on
  {Artificial} {Intelligence} and {Statistics}}, pages 153--160. PMLR, April
  2009.
\newblock URL \url{https://proceedings.mlr.press/v5/erhan09a.html}.

\bibitem[Brown et~al.(2020)Brown, Mann, Ryder, Subbiah, Kaplan, Dhariwal,
  Neelakantan, Shyam, Sastry, Askell, Agarwal, Herbert-Voss, Krueger, Henighan,
  Child, Ramesh, Ziegler, Wu, Winter, Hesse, Chen, Sigler, Litwin, Gray, Chess,
  Clark, Berner, McCandlish, Radford, Sutskever, and
  Amodei]{brown_language_2020}
Tom Brown, Benjamin Mann, Nick Ryder, Melanie Subbiah, Jared~D Kaplan, Prafulla
  Dhariwal, Arvind Neelakantan, Pranav Shyam, Girish Sastry, Amanda Askell,
  Sandhini Agarwal, Ariel Herbert-Voss, Gretchen Krueger, Tom Henighan, Rewon
  Child, Aditya Ramesh, Daniel Ziegler, Jeffrey Wu, Clemens Winter, Chris
  Hesse, Mark Chen, Eric Sigler, Mateusz Litwin, Scott Gray, Benjamin Chess,
  Jack Clark, Christopher Berner, Sam McCandlish, Alec Radford, Ilya Sutskever,
  and Dario Amodei.
\newblock Language models are few-shot learners.
\newblock In H.~Larochelle, M.~Ranzato, R.~Hadsell, M.F. Balcan, and H.~Lin,
  editors, \emph{Advances in Neural Information Processing Systems}, volume~33,
  pages 1877--1901. Curran Associates, Inc., 2020.
\newblock URL
  \url{https://proceedings.neurips.cc/paper_files/paper/2020/file/1457c0d6bfcb4967418bfb8ac142f64a-Paper.pdf}.

\bibitem[McCloskey and Cohen(1989)]{mccloskey_catastrophic_1989}
Michael McCloskey and Neal~J. Cohen.
\newblock Catastrophic {Interference} in {Connectionist} {Networks}: {The}
  {Sequential} {Learning} {Problem}.
\newblock \emph{Psychology of Learning and Motivation - Advances in Research
  and Theory}, 24\penalty0 (C):\penalty0 109--165, January 1989.
\newblock ISSN 0079-7421.
\newblock \doi{10.1016/S0079-7421(08)60536-8}.
\newblock URL
  \url{http://www.scopus.com/inward/record.url?scp=77957064197&partnerID=8YFLogxK}.

\bibitem[Smith and Topin(2019)]{smith_super-convergence_2018}
Leslie~N. Smith and Nicholay Topin.
\newblock {Super-convergence: very fast training of neural networks using large
  learning rates}.
\newblock In \emph{Artificial Intelligence and Machine Learning for
  Multi-Domain Operations Applications}, volume 11006, page 1100612.
  International Society for Optics and Photonics, SPIE, 2019.
\newblock \doi{10.1117/12.2520589}.

\bibitem[Loshchilov and Hutter(2017)]{loshchilov_sgdr_2017}
Ilya Loshchilov and Frank Hutter.
\newblock {SGDR}: Stochastic gradient descent with warm restarts.
\newblock In \emph{International Conference on Learning Representations}, 2017.

\bibitem[Smith(2017)]{smith_cyclical_2017}
Leslie~N. Smith.
\newblock Cyclical {Learning} {Rates} for {Training} {Neural} {Networks}.
\newblock \emph{arXiv:1506.01186 [cs]}, April 2017.
\newblock URL \url{http://arxiv.org/abs/1506.01186}.
\newblock arXiv: 1506.01186.

\bibitem[Chantry et~al.(2021)Chantry, Christensen, Dueben, and
  Palmer]{chantry_opportunities_2021}
Matthew Chantry, Hannah Christensen, Peter Dueben, and Tim Palmer.
\newblock Opportunities and challenges for machine learning in weather and
  climate modelling: hard, medium and soft {AI}.
\newblock \emph{Philosophical Transactions of the Royal Society A:
  Mathematical, Physical and Engineering Sciences}, 379\penalty0
  (2194):\penalty0 20200083, April 2021.
\newblock \doi{10.1098/rsta.2020.0083}.
\newblock URL
  \url{https://royalsocietypublishing.org/doi/10.1098/rsta.2020.0083}.
\newblock Publisher: Royal Society.

\bibitem[Gilleland et~al.(2009)Gilleland, Ahijevych, Brown, Casati, and
  Ebert]{gilleland_intercomparison_2009}
Eric Gilleland, David Ahijevych, Barbara~G. Brown, Barbara Casati, and
  Elizabeth~E. Ebert.
\newblock Intercomparison of {Spatial} {Forecast} {Verification} {Methods}.
\newblock \emph{Weather and Forecasting}, 24\penalty0 (5):\penalty0 1416--1430,
  October 2009.
\newblock ISSN 1520-0434, 0882-8156.
\newblock \doi{10.1175/2009WAF2222269.1}.
\newblock URL
  \url{https://journals.ametsoc.org/view/journals/wefo/24/5/2009waf2222269_1.xml}.
\newblock Publisher: American Meteorological Society Section: Weather and
  Forecasting.

\bibitem[Zhang et~al.(2017)Zhang, Zheng, and Qi]{zhang_deep_2017}
Junbo Zhang, Yu~Zheng, and Dekang Qi.
\newblock Deep {Spatio}-{Temporal} {Residual} {Networks} for {Citywide} {Crowd}
  {Flows} {Prediction}.
\newblock \emph{arXiv:1610.00081 [cs]}, January 2017.
\newblock URL \url{http://arxiv.org/abs/1610.00081}.
\newblock arXiv: 1610.00081.

\bibitem[Mathieu et~al.(2016)Mathieu, Couprie, and LeCun]{mathieu_deep_2016}
Micha{\"{e}}l Mathieu, Camille Couprie, and Yann LeCun.
\newblock Deep multi-scale video prediction beyond mean square error.
\newblock In Yoshua Bengio and Yann LeCun, editors, \emph{International
  Conference on Learning Representations}, 2016.

\bibitem[Hess and Boers(2022)]{hess_deep_2022}
Philipp Hess and Niklas Boers.
\newblock Deep {Learning} for {Improving} {Numerical} {Weather} {Prediction} of
  {Heavy} {Rainfall}.
\newblock \emph{Journal of Advances in Modeling Earth Systems}, 14\penalty0
  (3):\penalty0 e2021MS002765, 2022.
\newblock ISSN 1942-2466.
\newblock \doi{10.1029/2021MS002765}.
\newblock URL
  \url{https://onlinelibrary.wiley.com/doi/abs/10.1029/2021MS002765}.

\bibitem[Kaplan et~al.(2020)Kaplan, McCandlish, Henighan, Brown, Chess, Child,
  Gray, Radford, Wu, and Amodei]{kaplan_scaling_2020}
Jared Kaplan, Sam McCandlish, Tom Henighan, Tom~B. Brown, Benjamin Chess, Rewon
  Child, Scott Gray, Alec Radford, Jeffrey Wu, and Dario Amodei.
\newblock Scaling {Laws} for {Neural} {Language} {Models}, January 2020.
\newblock URL \url{http://arxiv.org/abs/2001.08361}.
\newblock arXiv:2001.08361 [cs, stat].

\bibitem[Palmer(2017)]{palmer_primacy_2017}
Tim Palmer.
\newblock The primacy of doubt: {Evolution} of numerical weather prediction
  from determinism to probability.
\newblock \emph{Journal of Advances in Modeling Earth Systems}, 9\penalty0
  (2):\penalty0 730--734, 2017.
\newblock ISSN 1942-2466.
\newblock \doi{10.1002/2017MS000999}.

\bibitem[Kashinath et~al.(2021)Kashinath, Mustafa, Albert, Wu, Jiang,
  Esmaeilzadeh, Azizzadenesheli, Wang, Chattopadhyay, Singh, Manepalli,
  Chirila, Yu, Walters, White, Xiao, Tchelepi, Marcus, Anandkumar, Hassanzadeh,
  and Prabhat]{kashinath_physics-informed_2021}
K.~Kashinath, M.~Mustafa, A.~Albert, J-L. Wu, C.~Jiang, S.~Esmaeilzadeh,
  K.~Azizzadenesheli, R.~Wang, A.~Chattopadhyay, A.~Singh, A.~Manepalli,
  D.~Chirila, R.~Yu, R.~Walters, B.~White, H.~Xiao, H.~A. Tchelepi, P.~Marcus,
  A.~Anandkumar, P.~Hassanzadeh, and null Prabhat.
\newblock Physics-informed machine learning: case studies for weather and
  climate modelling.
\newblock \emph{Philosophical Transactions of the Royal Society A:
  Mathematical, Physical and Engineering Sciences}, 379\penalty0
  (2194):\penalty0 20200093, April 2021.
\newblock \doi{10.1098/rsta.2020.0093}.
\newblock URL
  \url{https://royalsocietypublishing.org/doi/full/10.1098/rsta.2020.0093}.

\bibitem[Cuomo et~al.(2022)Cuomo, Di~Cola, Giampaolo, Rozza, Raissi, and
  Piccialli]{cuomo_2022}
Salvatore Cuomo, Vincenzo~Schiano Di~Cola, Fabio Giampaolo, Gianluigi Rozza,
  Maziar Raissi, and Francesco Piccialli.
\newblock Scientific machine learning through physics--informed neural
  networks: Where we are and what's next.
\newblock \emph{Journal of Scientific Computing}, 92\penalty0 (3):\penalty0 88,
  2022.
\newblock \doi{10.1007/s10915-022-01939-z}.
\newblock URL \url{https://doi.org/10.1007/s10915-022-01939-z}.

\bibitem[Alet et~al.(2019)Alet, Jeewajee, Villalonga, Rodriguez, Lozano-Perez,
  and Kaelbling]{alet_graph_2019}
Ferran Alet, Adarsh~Keshav Jeewajee, Maria~Bauza Villalonga, Alberto Rodriguez,
  Tomas Lozano-Perez, and Leslie Kaelbling.
\newblock Graph element networks: adaptive, structured computation and memory.
\newblock In Kamalika Chaudhuri and Ruslan Salakhutdinov, editors,
  \emph{Proceedings of the 36th International Conference on Machine Learning},
  volume~97 of \emph{Proceedings of Machine Learning Research}, pages 212--222.
  PMLR, 09--15 Jun 2019.

\bibitem[Watt-Meyer et~al.(2023)Watt-Meyer, Dresdner, McGibbon, Clark, Henn,
  Duncan, Brenowitz, Kashinath, Pritchard, Bonev, Peters, and
  Bretherton]{watt-meyer_ace_2023}
Oliver Watt-Meyer, Gideon Dresdner, Jeremy McGibbon, Spencer~K. Clark, Brian
  Henn, James Duncan, Noah~D. Brenowitz, Karthik Kashinath, Michael~S.
  Pritchard, Boris Bonev, Matthew~E. Peters, and Christopher~S. Bretherton.
\newblock {ACE}: {A} fast, skillful learned global atmospheric model for
  climate prediction, December 2023.
\newblock URL \url{http://arxiv.org/abs/2310.02074}.
\newblock arXiv:2310.02074 [physics].

\bibitem[Liu et~al.(2024)Liu, Zhang, Li, Yan, Gao, Chen, Yuan, Huang, Sun, Gao,
  He, and Sun]{liu_sora_2024}
Yixin Liu, Kai Zhang, Yuan Li, Zhiling Yan, Chujie Gao, Ruoxi Chen, Zhengqing
  Yuan, Yue Huang, Hanchi Sun, Jianfeng Gao, Lifang He, and Lichao Sun.
\newblock Sora: {A} {Review} on {Background}, {Technology}, {Limitations}, and
  {Opportunities} of {Large} {Vision} {Models}, February 2024.
\newblock URL \url{http://arxiv.org/abs/2402.17177}.
\newblock arXiv:2402.17177 [cs].

\bibitem[Buizza et~al.(1999)Buizza, Milleer, and
  Palmer]{buizza_stochastic_1999}
Roberto Buizza, M~Milleer, and Tim~N Palmer.
\newblock Stochastic representation of model uncertainties in the {ECMWF}
  ensemble prediction system.
\newblock \emph{Quarterly Journal of the Royal Meteorological Society},
  125\penalty0 (560):\penalty0 2887--2908, 1999.
\newblock Publisher: Wiley Online Library.

\end{thebibliography}


\end{document}